%% file: sample-acmsmall.tex
\documentclass[acmsmall]{acmart}

\usepackage{acronym} 
\usepackage{algorithmic}
\usepackage{booktabs}
\usepackage{clrscode3e}
\usepackage{hyperref}
\usepackage{multirow}
\usepackage{graphicx}
\usepackage{subfigure}
\usepackage{textcomp}
\usepackage{xcolor}
\usepackage{tabularx}
\usepackage{xspace}
\usepackage{caption}
\usepackage[normalem]{ulem}
\usepackage{cleveref}
\usepackage{amsmath}
\usepackage{varwidth}
\usepackage{paralist}
\usepackage{wrapfig}
\usepackage{bm}
\usepackage{xcolor}
\usepackage[normalem]{ulem}
\usepackage{enumitem}
\usepackage{tikz}
\useunder{\uline}{\ul}{}
\usepackage[ruled, vlined, linesnumbered]{algorithm2e}
\usepackage{mathrsfs}
\usepackage{tabularx}
\usepackage{threeparttable}

\acrodef{FL}{Federated Learning}
\acrodef{PCDR}{Privacy-preserving Cross-domain Recommendation}
\acrodef{DPCSR}{Domain-level Privacy-preserving Cross-domain Sequential Recommendation}
\acrodef{FFMSR}{Filter-enhanced Federated Multi-layer Semantic Representation}
\acrodef{PFCR}{Prompt-enhanced Federated Content Representation}
\acrodef{PLMs}{Pre-trained Language Models}
\acrodef{PLM}{Pre-trained Language Model}
\acrodef{MoE}{Mixture of Experts}
\acrodef{PW}{Parameter Whitening}
\acrodef{MLP}{Multi Layer Perceptron}
\acrodef{FFT}{Fast Fourier Transform}
\acrodef{CDR}{Cross-Domain Recommendation}
\acrodef{LDP}{Local Differential Privacy}
\acrodef{GDPR}{General Data Protection Regulation}
\acrodef{PPCDR}{Privacy-Preserving Cross-Domain Recommendation}
\acrodef{LLMs}{Large Language Models}
\acrodef{PQ}{Product Quantization}
\AtBeginDocument{%
  \providecommand\BibTeX{{%
    \normalfont B\kern-0.5em{\scshape i\kern-0.25em b}\kern-0.8em\TeX}}}

\setcopyright{acmcopyright}
\copyrightyear{2023}
\acmYear{xxx}
\acmDOI{10.1145/0000000.000000}

\acmJournal{TOIS}
\acmVolume{0}
\acmNumber{0}
\acmArticle{0}
\acmMonth{0}

\begin{document}

\title{Federated Semantic Learning for Privacy-preserving Cross-domain Recommendation}

\author{Ziang Lu}
\email{zianglu@outlook.com}
\affiliation{
  \institution{Shandong Normal University}
  \city{Jinan}
  \state{Shandong}
  \country{China}
  \postcode{250358}
}
\author{Lei Guo}
\authornote{Corresponding Author.}
\email{leiguo.cs@gmail.com}
\affiliation{
  \institution{Shandong Normal University}
  \city{Jinan}
  \state{Shandong}
  \country{China}
  \postcode{250358}
}
\author{Xu Yu}
\email{yuxu0532@upc.edu.cn}
\affiliation{
  \institution{China University of Petroleum (East China)}
  \city{Qingdao}
  \state{Shandong}
  \country{China}
  \postcode{266580}
}
\author{Zhiyong Cheng}
\email{jason.zy.cheng@gmail.com}
\affiliation{
  \institution{Hefei University of Technology}
  \city{Hefei}
  \state{Anhui}
  \country{China}
  \postcode{230009}
}
\author{Xiaohui Han}
\email{xiaohhan@gmail.com}
\affiliation{
  \institution{Key Laboratory of Computing Power Network and Information Security, Ministry of Education, Qilu University of Technology (Shandong Academy of Sciences)}
  \city{Jinan}
  \state{Shandong}
  \country{China}
  \postcode{250014}
}
\author{Lei Zhu}
\email{leizhu0608@gmail.com}
\affiliation{
  \institution{Tongji University}
  \city{Jinan}
  \state{Shanghai}
  \country{China}
  \postcode{200092}
}



\renewcommand{\shortauthors}{Lu, and Guo et al.}

\begin{abstract}
In the evolving landscape of recommender systems, the challenge of effectively 
conducting privacy-preserving \ac{CDR}, especially under strict non-overlapping constraints, has emerged as a key focus.
Despite extensive research has made significant progress, several limitations still exist: 1) 
Previous semantic-based methods fail to deeply exploit rich textual information, since they quantize the text into codes, losing its original rich semantics.
2) The current solution solely relies on the text-modality, while the synergistic effects with the ID-modality are ignored.
3) Existing studies
do not consider the impact of irrelevant semantic features, leading to inaccurate semantic representation.
To address these challenges, we introduce federated semantic learning and devise FFMSR as our solution.
For Limitation 1, we locally learn items' semantic encodings from their original texts by a multi-layer semantic encoder, and then cluster them on the server to facilitate the transfer of semantic knowledge between domains.
To tackle Limitation 2, we integrate both ID and Text modalities on the clients, and utilize them to learn different aspects of items.
To handle Limitation 3, a \ac{FFT}-based filter and a gating mechanism are developed to alleviate the impact of irrelevant semantic information in the local model.
We conduct extensive experiments on two real-world datasets, and the results demonstrate the superiority of our FFMSR method over other SOTA methods. Our source codes are publicly available at: \textcolor{blue}{\url{https://github.com/Sapphire-star/FFMSR}}.
\end{abstract}



\maketitle

\input{01-introductin}
\input{02-realtedwork}

\input{03-proposedmethod}
\input{04-setup}
\input{05-result}
\input{06-analysis}
\input{07-conclusions}


\begin{acks}
This work was supported by the National Natural Science Foundation of China (No. 62372277), Natural Science Foundation of Shandong Province (Nos. ZR2022MF257, ZR2020YQ47), Computing Power Internet and Information Security Key Laboratory of Ministry of Education (No. 2023ZD024).
\end{acks}

\bibliographystyle{ACM-Reference-Format}
\bibliography{references}

\end{document}

%% file: 01-introductin.tex
\section{Introduction}
In recent years, with the proliferation of various platform services, how to effectively utilize knowledge from other platforms to enhance recommendation performance has become a key research issue in the field of recommender systems.
Methods for \acf{CDR} have received extensive attention from researchers in recent years~\cite{xie2022contrastive, zhao2023cross, guo2023disentangled, liu2024inter, guo2021gcn}.
\ac{CDR} is the task that aims to enhance recommendation performance in the target domain by extracting valuable information from auxiliary domains, particularly important in cases where the target domain suffers from data sparsity.
Although recent \ac{CDR} methods have shown promising results~\cite{liu2024mcrpl, liu2023joint, guo2023dan, du2023distributional, guo2023automated}, they often fail in the privacy-preserving scenarios, since the leakage of domain and user privacy.
Strict data protection regulations, such as the \ac{GDPR}~\cite{voigt2017eu}, prohibit the sharing of sensitive personal information across platforms, including user IDs, age, gender, reviews, and interaction data. Such legal restrictions directly block the sharing of inter-domain data, rendering traditional \ac{CDR} methods inoperable in privacy-preserving environments. For example, TikTok and Bilibili are two popular video-watching platforms in China. They have accumulated different types of user behavior patterns, if they want to learn from each other to further enhance user experiences, they must share user data for model training in traditional recommendation methods.
However, due to corporate confidentiality and relevant regulatory restrictions, these two platforms cannot share their data freely.

Faced with the above challenge, researchers have begun exploring new strategies for \ac{PPCDR}. 
Typical methods for \ac{PPCDR} often adopt federated learning to allow multiple organizations to collaboratively train models without data exchange. 
For instance, Zhang et al.~\cite{zhang2024feddcsr} propose a federated learning framework that employs disentangled representation learning to separate each domain’s model into domain-shared and domain-exclusive parts, only aggregating the domain-shared component at the server and enhancing the learning of domain-exclusive features through contrastive learning tasks. 
Zhao et al.~\cite{zhao2024personalized} first use federated learning to learn a common preference bridge for all users, and then utilize a meta-learning network to generate personalized preference bridges for each user on this basis.

However, most previous \ac{PPCDR} studies mainly rely on overlapping users between domains as a bridge to perform information transfer.
It is unrealistic to assume that one domain (or platform) knows all or part of the shared users from another domain because the user's identity information is protected.
To our knowledge, only one pioneer work, i.e., PFCR~\cite{guo2024prompt}, considers \ac{PPCDR} under the non-overlapping constraint.
PFCR~\cite{guo2024prompt} employs vector quantization techniques to model the semantic similarity of item description texts across different domains and devise a federated content representation learning method for domain knowledge transfer.
But it only learns the item contents in a shallow manner and exploits the text-modality to perform recommendations, which solely reaches sub-optimal results.
Specifically, PFCR still has the following limitations: 1) It has severely lost the rich semantic information contained in the original texts due to the user of \ac{PQ}. The PQ method transforms the continuous representation of the item text into multiple discrete codes, which can only reflect the approximate semantics within each subspace, and is unable to precisely capture the fine-grained meanings of items' text. Furthermore, the characteristics of these codes largely depend on the updates of the code embedding table. As the model continues to update, changes in this table may cause the features to gradually deviate from the original text, exacerbating the loss of semantic information. The shallow learning mode of semantics prevents us from discovering more common information between domains.
2) It only leverages the text-modal to derive the code embeddings while the collaborative filtering signals embedded in the ID-modal are ignored. 
ID-modality is one of the most effective signals and has been proved in traditional recommendation methods~\cite{kang2018self, sun2019bert4rec, he2020lightgcn, wang2019neural}.
Recent studies find that simultaneously considering these two modalities can capture item information more comprehensively since they can model different aspects of items~\cite{xu2024sequence, cheng2024empowering, wang2024aligned, zhang2024disentangling}.
3) It fails to filter out irrelevant semantic information obtained through the federated learning in both text- and ID- modalities, leading to inaccurate semantic representation.

To address the above challenges, we devise a federated semantic learning framework for \ac{PPCDR} and propose a \ac{FFMSR} as our solution. Our method operates in a fully non-overlapping federated cross-domain scenario where there is no overlap between users and items across different domains, and our primary idea is to facilitate the exchange of information between the well-learned semantic encodings across domains.
Specifically, to deal with Limitation 1, we process all item description texts through a \ac{PLM} to preserve multi-layer semantic encodings in each client and then transform them into semantic embeddings using a \ac{MoE} adapter. In the subsequent federated training phase, each client uploads the encodings that aggregate different layers of semantics to the server.
Thereafter, the server clusters them to facilitate the semantic transfer between domains.
To tackle Limitation 2, we utilize the ID modality as the supervisory signal for the fusion guidance of multi-layer semantic embeddings. The ID modality can precisely extract the most valuable information for modeling user-item relationships from the multi-layer semantic representations.
The item representations in two modalities are then combined together and fed to a sequence encoder for further modeling.
To handle Limitation 3, 
we design a gating mechanism to dynamically allocate the contribution of ID-modality. 
To filter out irrelevant semantic meanings in the clustering embedding from the server, a FFT-based filter layer with a learnable filter parameter is adopted. It can more effectively identify and eliminate noise at specific frequencies through frequency domain analysis without destroying useful information in the original signal.
To enhance the capabilities of the followed sequence encoder in extracting useful information from the mixed item representations, we also incorporate the filter layer into it. Specifically, we add the FFT-based filter layers before the self-attention layers to process complex input features. We also use residual connections and layer normalization after the filter layer to avoid training instability and to better integrate with the Transformer structure.

Our contributions can be summarized as follows:
\begin{itemize}
    \item We propose a semantic-enhanced federated learning framework \ac{FFMSR} for \ac{PPCDR}, enabling federated training in completely non-overlapping and privacy-preserving scenarios with deeply encoded semantic information.  
    \item We use a multi-layer \ac{PLM} and MoE Adapter to model different levels of semantics from items' original description texts to prevent semantic loss caused by \ac{PQ} method.
    \item We integrate ID-modality to text-modality by letting it participate in the fusion block on the semantic encodings of items' texts, enabling us to model different aspects of items.
    \item We devise a \ac{FFT}-based filter layer to filter out the noises within the clustered embeddings and hybrid sequence embeddings. To identify the irrelevant ID features, a gating mechanism is applied.  
    \item We conduct experiments on two real-world datasets and the experimental results demonstrate the superiority of our proposed \ac{FFMSR} method compared with the recent SOTA methods.    
\end{itemize}

%% file: 02-realtedwork.tex
\section{Related Work}
This section takes the privacy-preserving \ac{CDR}, semantic-enhanced recommendation, and filter-enhanced recommendation as related works.
\subsection{Privacy-preserving Cross-domain Recommendation}
In modern recommender systems, users often interact with products across multiple platforms. Privacy-preserving \ac{CDR} is the task that aims at effectively transferring knowledge between different domains while avoiding the disclosure of user privacy-related data. According to how privacy protection is achieved, these methods can be divided into two categories: federated-based~\cite{yan2022fedcdr, meihan2022fedcdr, zhang2024feddcsr, zhang2024fedhcdr, zhang2021vertical, wan2023fedpdd, chen2023win, kalloori2021horizontal, cai2022privacy, zhao2024personalized, tian2023privacy} and knowledge transfer-based transferring approaches~\cite{wang2024privacy, chen2022differential, liu2023differentially, liao2023ppgencdr}.

Federated-based studies often rely on a user or an organization acting as a client to transmit parameters to an intermediate server to facilitate information integration across domains. For instance, 
Tian et al.~\cite{tian2023privacy} treat each domain as a client, keeping user interaction data locally on each client. They design a GNN to model global preferences across all domains and local preferences within specific domains. Each client participates in bidirectional message exchange and propagation during the federated interaction process. After receiving the global user preferences, clients perform personalized aggregation, encrypt the data, and then share it with other clients. Wu et al.~\cite{meihan2022fedcdr} devise a federated \ac{CDR} framework by treating users as clients, where individual modules are developed to learn each user's personalized preferences and a transfer module is learned to facilitate knowledge transfer between domains. In this method, clients only transmit parameters of non-personal modules to the server. 

Knowledge transfer-based methods typically involve the protection and transfer of user-related parameters. For example, Chen et al.~\cite{chen2022differential} map the user co-occurrence matrix from a high-dimensional space to a low-dimensional space while preserving the geometric similarity between users. They also propose a cross-domain model based on deep autoencoders and DNN to model both the source and target rating matrices, aiming to maintain data privacy and effectiveness in the recommendation. Liao et al.~\cite{liao2023ppgencdr} design a privacy-preserving CDR framework consisting primarily of a privacy protection generator module in the source domain and a cross-domain recommendation module in the target domain. The privacy protection module is used to generate privacy-protected user parameters, while the cross-domain recommendation module extracts user preference features from the generator and combines them with target domain user features to enhance the recommendation performance in the target domain. 
However, most privacy-preserving CDR methods rely on overlapping users between domains for knowledge transfer. The only method that considers non-overlapping scenarios, PFCR~\cite{guo2024prompt}, neglects the richness of semantic information.
\subsection{Semantic-enhanced Recommendation}
Semantic information serves as a critical enhancement to recommendation algorithms in contemporary recommender systems.   The utilization of semantic representations within these models categorizes their applications into two distinct types: Text-only~\cite{hou2022towards, hou2023learning, mu2022id, shin2021one4all, li2023text, peng2023towards} and Text-ID based methods~\cite{cheng2024empowering, li2024enhancing, xu2024sequence, zhang2024id}. 
The Text-only approach utilizes semantic vectors derived solely from the textual descriptions of items as their representation vectors. For example, Hou et al.~\cite{hou2022towards} employ \ac{PLMs} to encode the descriptions of items and then process these through a \ac{MoE} layer to learn a generic text representation that can be transferred across different contexts. Additionally, they introduce two types of contrastive learning tasks to acquire a more generalizable sequential representation. 

The Text-ID approach in recommendation models leverages both the ID and textual information of items. This method integrates these data sources collaboratively or through fusion, aiming to more effectively utilize the knowledge contained within different modalities. For instance, Cheng et al.~\cite{cheng2024empowering} propose an end-to-end dual-stream framework to learn user preferences from both ID and text modalities. They incorporated two contrastive learning tasks to align representations across modalities and employed a dual-stream encoder to adjust the relationships between the modalities dynamically. Xu et al.~\cite{xu2024sequence} introduce a novel Text-ID semantic fusion method. In their approach, representations from both modalities are transformed to the frequency domain using the \ac{FFT}. They then perform element-wise multiplication of these transformed representations before converting them back to the time domain to obtain a fused item representation. These semantic-enhanced recommendation methods are primarily designed for single-domain recommendation tasks. Their models lack specific configurations that would enable cross-domain functionality.

It is worth emphasizing that to avoid the issues proposed in~\cite{hou2023learning} regarding the "text $\rightarrow$ representation" paradigm, we address them in the following ways: 1) To avoid the model over-emphasize the effect of text features, we introduce the ID modality to model users' collaborative filtering interests and interactive behaviors. Specifically, in the fusion block of the Semantic Extraction module, we use items' ID modality as collaborative filtering signals to guide the dynamic allocation process on the semantic encodings from multiple PLM layers. Moreover, in the Semantic Filtering module, we further fuse both the ID and text modalities as the foundation of the sequence encoding to ensure the following sequence encoder can model both users' collaborative interactions and items' descriptive texts. 2) To address the domain gap challenge, we devise an improved clustering method on the server side and a FFT-based filter layer in the client. Compared to the simple average aggregation method used in the traditional K-means algorithm, we design a weighted aggregation strategy based on the distance between each data point and its corresponding cluster centroid.
Our design allows each cluster to focus more on semantically relevant features, effectively reducing the negative impact of domain gaps and improving the semantic consistency and accuracy of clustering results. 
The FFT-based filter layer is devised for further filtering out the negative noise caused by the domain gap. It can more effectively identify and eliminate noise at specific frequencies through frequency domain analysis without destroying useful information in the original signal.
\subsection{Denoising in Recommendation}
\label{sec:denoise}
As user interactions with various platforms and products increase, implicit feedback derived from these interactions may contain noise, making it difficult to accurately model and predict user interests. Current denoising approaches often involve correcting potentially noisy items directly at the original sequence level~\cite{gao2022self, wang2021denoising, chen2022denoising, sun2021does}. For instance,  Wang et al.~\cite{wang2021denoising} propose an adaptive denoising strategy comprising two main training methods: truncated loss, which truncates the loss from hard interactions, and reweighted loss, which dynamically assigns smaller weights to hard interactions. Gao et al.~\cite{gao2022self} design a self-guided denoising paradigm, where they collect memory interaction data during the initial training phase. This data is then used to guide the denoising of implicit feedback in subsequent training phases using a meta-learning strategy. Additionally, some advanced works suggest filtering noise directly at the sequence representation vector level through learnable modules~\cite{zhou2022filter, ai2022fourier, zhang2022attention, du2023frequency, zhang2023contrastivet, zhang2023contrastive}. For example, Zhou et al.~\cite{zhou2022filter} introduce frequency domain denoising methods commonly used in signal processing into recommendation models. They first transform the input representation vectors to the frequency domain using \ac{FFT}, filter the noise in the frequency domain with a learnable filter parameter, and finally use inverse \ac{FFT} to transform the filtered representations back to the time domain. The \ac{FFT}-based method can better capture the global information of sequences and is simpler and more efficient compared to methods that correct noisy items.

\subsection{Differences}
In comparison to the pioneering work PFCR~\cite{guo2024prompt}, which is also oriented towards domain-level privacy-preserving cross-domain sequential recommendation task, PFCR employs vector quantization technology to model the semantic similarity of item texts across different domains, and conducts cross-domain knowledge transfer based on the code embeddings used to represent items. In FFMSR, we leverage the clustering method on the server to achieve cross-domain semantic knowledge transfer. Our approach can avoid the loss of rich semantic information that comes with vector quantization while improving the model's performance.

Compared with other recommendation methods that use text as auxiliary information, although it is a common practice to obtain semantic encoding through PLM and then transform it through the MoE adapter, as seen in UniSRec~\cite{hou2022towards}, we do not directly leverage the existing MoE adapter structure in the semantic extraction module. To take into account the rich semantics contained in the semantic encodings, we apply a multi-layer semantic encoding method to model different levels of semantic information. Furthermore, to simultaneously consider the ID modality to obtain users' collaborative filtering interests, we then leverage the ID embeddings as a supervisory signal to guide the semantic fusion process.

In contrast to conventional approaches that filter noise at the level of the raw sequence~\cite{gao2022self, wang2021denoising, chen2022denoising, sun2021does}, we employ a \ac{FFT}-based filter to mitigate adverse noise at the representation level across various modalities. Furthermore, in contrast to other works that leverage FFT~\cite{zhou2022filter, ai2022fourier, zhang2022attention, du2023frequency, zhang2023contrastivet, zhang2023contrastive}, we integrate an FFT-based filter layer into a Transformer-style sequence encoder, thereby enhancing its ability to extract information from a mixture of multiple modal representations.

%% file: 03-proposedmethod.tex
\begin{figure}
    \centering    \includegraphics[width=8cm]{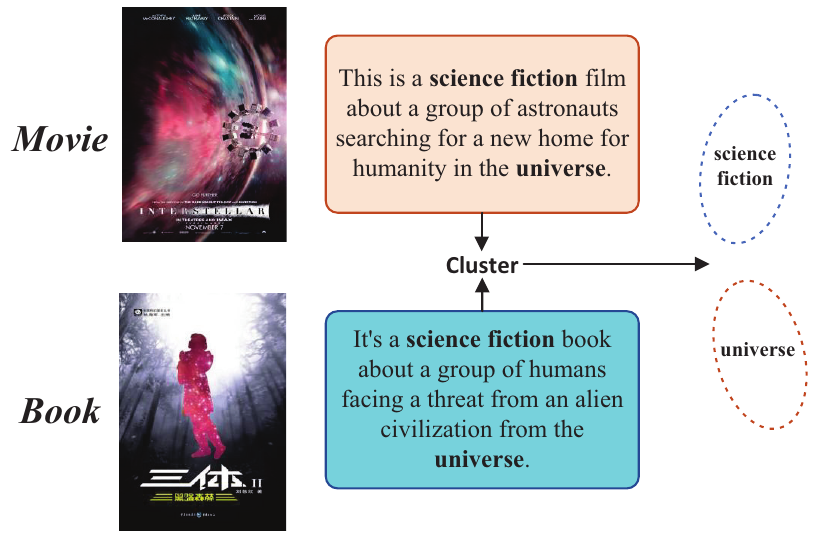}
    \caption{An example of capturing common semantic information between different items through clustering.}
    \label{fig:example}
\end{figure}

\section{Methodology}
This section begins by presenting the preliminaries and the overarching framework of our design, followed by a comprehensive explanation of our implementation approach.

\subsection{Preliminaries}
Consider two domains/platforms, A and B (we will treat domain and platform as equivalent concepts). We assume that these domains have entirely non-overlapping users and items. We make this assumption because in real applications, user identity information is commercial privacy data. Even if some users interact with items in both domains at the same time, the user identity and interaction information are not allowed to be leaked to the outside within the organization.
We denote the user sets of these two domains as $\mathcal{U}^A$ and $\mathcal{U}^B$, the item sets as $\mathcal{A}=\{A_1, A_2,\ldots, A_{M_A}\}$ and $\mathcal{B}=\{B_1, B_2,\ldots, B_{M_B}\}$, where $M_A$ and $M_B$ correspond to the number of items in domains A and B. Each item has a unique ID and corresponding description texts. Take domain A as an example, the description text of item  $A_i$ is denoted as $\mathcal{T}_i = \{ w_1, w_2,\ldots, w_i, \ldots, w_c\}$, where $w_i$ denotes the $i$-th word, and $c$ is the word number of $\mathcal{T}_i$. The interaction sequence of user ${u_i^A} \in \mathcal{U}^A$ to items in $\mathcal{A}$ can be expressed as $\mathcal{S}_i^A = \{ A_1, A_2,\ldots, A_j,\ldots\, A_{m_i}\}$, where $m_i$ denotes the length of $\mathcal{S}_i^A$. 

Our target is to utilize both textual information and unique IDs to accurately predict the next item that a user is most likely to interact with under the privacy-preserving constraint. The prediction probability is denoted as $P({A_{{m_i}+1}}|{\{ A_1, A_2,\ldots\, A_{m_i}\}})$.
Specifically, our task is to build a dual-target mapping function $f(\mathcal{x})$ for both domains and predict the probability of the candidate items in $\mathcal{A}$ (or $\mathcal{B}$) by exploring the cross-domain information under the non-overlapping and privacy-preserving constraints.
\begin{table}[htbp]
  \centering
  \caption{Summary of the main symbols and notations utilized in this work.}
  \begin{tabular}{ll}
    \toprule
    \textbf{Symbol} & \textbf{Notation} \\
    \midrule
    $\mathcal{U}^A$ & User set of domain A \\
    $u_i^A$ & User $i$ from domain A, $u_i^A \in \mathcal{U}^A$ \\
    $\mathcal{A}$ & Item set of domain A $\{A_1, A_2,\ldots, A_{M_A}\}$ \\
    $M_A$ & The number of items in domain A \\
    $\mathcal{S}_i^A$ & Interaction sequence of user $u_i^A$, $\mathcal{S}_i^A = \{ A_1, A_2,\ldots, A_j,\ldots\, A_{m_i}\}$ \\
    $m_i$ & Total number of items $u_i^A$ has interacted with \\
    $w$ & A word \\
    $\mathcal{T}_i$ & The description text of item $A_i$ in domain A, $\mathcal{T}_i = \{ w_i, w_2,\ldots, w_c\}$ \\
    $\bm{x}_i^L$ & Multi-layer semantic encoding of item $i$, $\bm{x}_i^L = \{\bm{x}_i^{L=n-a}, \bm{x}_i^{L=n-a+1}, \ldots, \bm{x}_i^{L=n}\}$ \\
    $n$ & Total number of layers in PLMs \\
    $a$ & The number of layers we choose to keep \\
    $\bm{T}^{L=j}$ & Semantic embedding table of $j$ layer \\
    $\bm{T}$ & Mixed-layer semantic embedding table \\
    $\bm{T}_i^A$ & Semantic embedding vector of interaction $\mathcal{S}_i^A$ \\
    $\bm{E}$ & ID embedding table \\
    $\bm{E}_i^A$ & ID embedding vector of interaction $\mathcal{S}_i^A$ \\
    $\bm{h}_i^A$ & User representation vector of user $u_i^A$ \\
    $K$ & The number of centroids in clustering \\
    \bottomrule
  \end{tabular}
  \label{tab:symbol}
\end{table}
\subsection{Overview of FFMSR}
\textbf{Motivation.}
To conduct privacy-preserving cross-domain recommendations,  we resort to the federated learning framework, where each domain acts as a client and users' interactions within the current domain will not be shared with others. The only public information is the description texts of the items.
To consider the non-overlapping constraint,  we naturally consider 
the universality of natural language and take the semantic meaning of items as a bridge to connect domains. Unlike the pioneer work PFCR~\cite{guo2024prompt}, we achieve this by deep semantic mining and clustering, enabling us to facilitate inter-domain semantic knowledge transfer and enhance the relevance between items. 
As illustrated in Fig~\ref{fig:example}, we consider two items from the movie and book domains. Suppose their themes both contain science fiction and universe, we can capture their commonalities and further strengthen these semantic features through cross-domain semantic clustering and aggregation, which can be further treated as an enhancement of domain semantic learning.


\textbf{Overall Framework.}
\begin{figure}
    \centering
    \includegraphics[width=13cm]{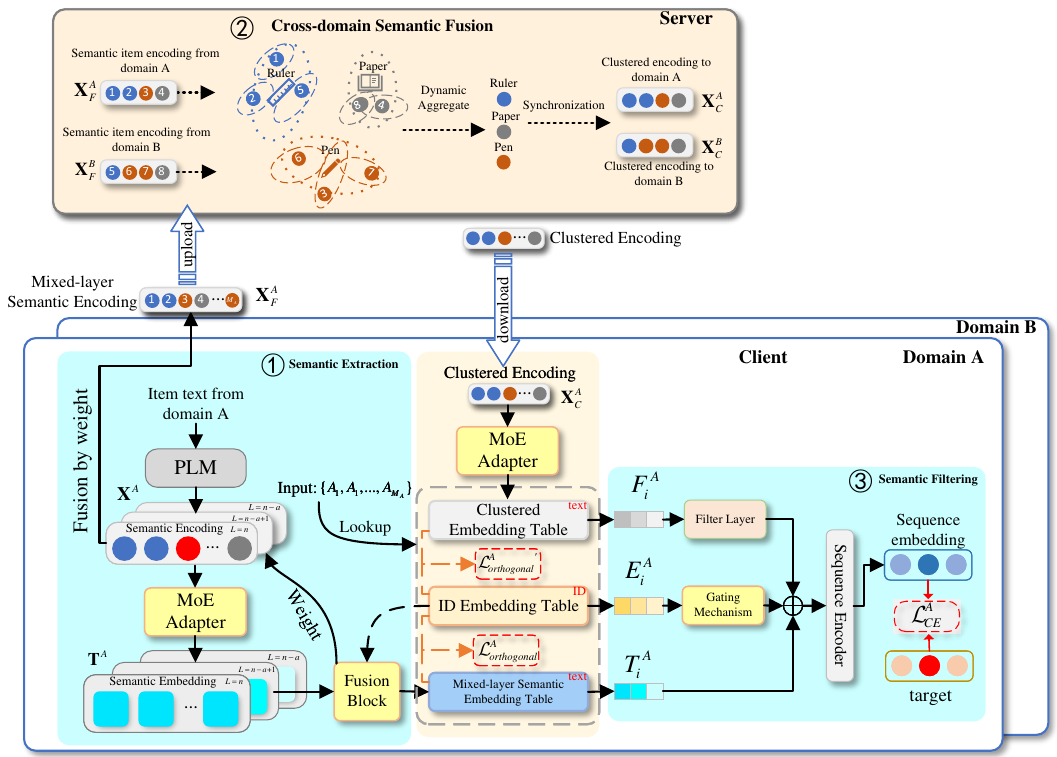}
    \caption{The system architecture of \ac{FFMSR}, which is mainly composed of semantic extraction, cross-domain semantic fusion, and semantic filtering components. (1) The semantic extraction module aims at learning the semantics of items' description texts at multi-levels of \ac{PLM}. The semantic encodings within it will be uploaded to the server for further fusion. (2) The cross-domain semantic fusion targets fusing the semantics in different domains by clustering. The clustered encoding will be sent back to the client for local updating.
    (3) The semantic Filtering intends to filter out the irrelevant semantics caused by semantic fusion and multi-modality utilization. }
    \label{fig:FFMSR}
\end{figure}
The system architecture of our \ac{FFMSR} method is depicted in Fig~\ref{fig:FFMSR}, 
which mainly consists of semantic extraction, cross-domain semantic fusion, and semantic filtering components.
1) In the semantic extraction component, we first obtain the semantic encodings of all items from their description texts by \ac{PLM} in its multiple layers. Then, we transform them into semantic embeddings by a MoE adapter, where an embedding table with multi-layer semantics is formed.
To consider the collaborative information that can be learned by atomic IDs, we integrate the ID-modality by treating them as signals to guide the fusion block to aggregate the semantic encodings in multiple layers. The output is named the mixed-layer semantic encoding.
Each client/domain uploads them to the server for further semantic fusion (the details can be seen in Section \ref{sec:semantic_extraction}).
2) In the cross-domain semantic fusion module, we cluster the uploaded semantic encodings for cross-domain semantic aggregation.
Our motivation for doing this is the universality of natural language, which enables us to connect disjoint domains.
Then, we utilize the centroid vector of its corresponding cluster as an item's semantic encoding.
Afterward, we send back the items' clustered encodings to all the clients, where the clustered embedding table is achieved after another \ac{MoE} adapter is applied to them (the details can be seen in Section \ref{sec:semantic_fusion}).
3) In the semantic filtering module, we first obtain three kinds of representations, i.e., the clustered embedding table, ID embedding table, and the mixed-layer semantic embedding table, for the items within the given sequence.
Since these representations are from mixed layers or modalities, they may contain irrelevant information to the target.
To obtain more accurate embeddings, we apply a \ac{FFT}-based filter layer and a gating mechanism to the clustered embeddings and ID embeddings, respectively.
Then, we fill the combined item vectors in a sequence encoder for sequence modeling.
Considering that the fusion of different types of item representations may have negative impacts, a filter-enhanced sequence encoder is then devised (the details can be seen in Section \ref{sec:semantic_filter}).

\subsection{Deep Semantic Extraction}\label{sec:semantic_extraction}
To deeply extract semantic information from the description texts of items, we do not follow the method in \cite{guo2024prompt} which uses generated quantized codes to model item semantics. Instead, we tend to learn the semantics directly from their original rich texts by a multi-layer \ac{PLM} followed by a \ac{MoE} adapter. To meet the privacy-preserving constraint, we only upload the semantic encodings that do not have connections with users' interactions to the server for cross-domain semantic fusion (the overall architecture can be seen in Figure. \ref{fig:FFMSR} (1)).

\subsubsection{Semantic Encoding by a Multi-layer \ac{PLM}}\label{sec:multi_layer_encoding}
To mine the deep semantic information contained in the items' description texts, we resort to the \acf{PLM}s for their text understanding capabilities, and use it as a text encoder to obtain the text encodings of items.
But unlike other text-based methods~\cite{hou2022towards, hou2023learning, yuan2023go} that only utilize the output of the final layer of \ac{PLM}s to represent items' semantics, we retain the output of multiple layers to keep more semantic information~\cite{vaswani2017attention}. It is worth emphasizing that we just take the \ac{PLM} as the text encoder to obtain the text encodings of items without any structural changes to it.
These encodings are the foundations of obtaining items' semantic presentations.

Motivated by ~\cite{ding2021zero}, for an item $i$, we first input the token $\texttt{[CLS]}$ at the beginning of its description text $\mathcal{T}_i=\{w_1, \ldots, w_c\}$, and then feed the word sequence into the pre-trained BERT~\cite{devlin2018bert} for semantic modeling. The resulting output of the $j$-th layer can be denoted as:
\begin{equation}
    \bm{x}_i^{L=j} = \text{BERT}^{L=j}(\left[\texttt{[CLS]}; w_1; \ldots; w_c\right]),
\end{equation}
where $[;]$ denotes the concatenation operation, $\bm{x}_i^{L=j} \in \mathbb{R}^{d_W}$ is the hidden vector of the token $\texttt{[CLS]}$ in the $j$-the layer, $d_W$ is the vector dimension. We keep the output of the last $a$ layers for item $i$ to serve as its initial multi-layer semantic encodings (denoted as $\boldsymbol{x}_i^L$), which is represented as:
\begin{equation}
     \bm{x}_i^L = \{\bm{x}_i^{L=n-a}, \bm{x}_i^{L=n-a+1}, \ldots, \bm{x}_i^{L=n}\},
\end{equation}
where $n$ is the total number of layers in BERT.
In experiments, we take $a$ as a hyper-parameter and evaluate it in Section~\ref{sec:hyper_param}.

\subsubsection{MoE Adapter.}\label{sec:MoE} Though BERT has the ability to encode texts into vectors, some studies have surfaced that the text encodings obtained from its direct output do not have enough discriminability between similar words~\cite{li2020sentence, huang2021whiteningbert}. 
To improve the distinguishability between semantic encoding and extract valuable semantic information~\cite{chen2022towards}, we then apply a \ac{MoE} adapter~\cite{hou2022towards} for further modeling. 

The \ac{MoE} adapter layer mainly consists of a \ac{PW} layer and a multi-expert gating mechanism.
Each expert within the \ac{MoE} framework is a \ac{PW} layer that aims to decorate the features of the input semantic encoding. For text encoding $\bm{x}_i$, after passing through the \ac{PW} layer, it is transformed into:
\begin{equation}
    \tilde{\bm{x}}_i^{L=j} = (\text{Dropout}(\bm{x}_i^{L=j}) - \mathbf{b}^{PW}) \cdot \mathbf{W}^{PW},
\end{equation}
where $\mathbf{W}^{PW} \in \mathbb{R}^{d_W \times d_V}$ and $\mathbf{b}^{PW} \in \mathbb{R}^{d_W}$ are learnable parameters, $d_V$ denotes the hidden size of the model. Using this approach, we reduce redundancy in text encodings, allowing the model to focus more on the most informative features and enhancing the distinctiveness of text features.

The gating mechanism in the \ac{MoE} Adapter is responsible for determining the contribution of each expert to the final semantic embedding table. This mechanism employs a set of trainable weights to dynamically allocate the input textual encoding to the most relevant experts based on the content. The gating weights are computed through a noisy gating process, which introduces stochasticity, encouraging exploration of different expert combinations and avoiding over-reliance on a single expert. The gating weights for each expert are computed as follows:
\begin{equation}
   \bm{g}_k= \text{Softmax}({\bm{x}_i^{L=j}}\mathbf{W}^{MoE} + \mathbf{\delta}), 
\end{equation}
where $\mathbf{W}^{MoE} \in \mathbb{R}^{{d_W} \times G}$ is a learnable parameter that dynamically adjusts the weights of each expert, $G$ is the number of experts. $\delta$ is the random Gaussian noise used to balance the expert loads.

The final output of the \ac{MoE} Adapter, representing the semantic embedding table of the input text, is a weighted combination of the outputs from all experts:
\begin{equation}
    \bm{t}_i^{L=j} = \sum_{k=1}^{G} \bm{g}_{k} \cdot \tilde{\bm{x}}_{i, k}^{L=j},
\end{equation}
where $\tilde{\bm{x}}_{i, k}^{L=j}$ is the ouput of the $k$-th \ac{PW} layer, $\bm{t}_i^{L=j} \in \mathbb{R}^{d_V}$ stands for the semantic vector of item $i$. We apply the \ac{MoE} adapter to each layer of semantic vectors for every item, resulting in each layer semantic embedding table, it can be expressed as:
\begin{equation}
    \bm{T}^{L=j} = \{\bm{t}_1^{L=j}, \bm{t}_2^{L=j}, \ldots, \bm{t}_M^{L=j}\}.
\end{equation}

\begin{figure}[t]
    \centering    \includegraphics[width=9cm]{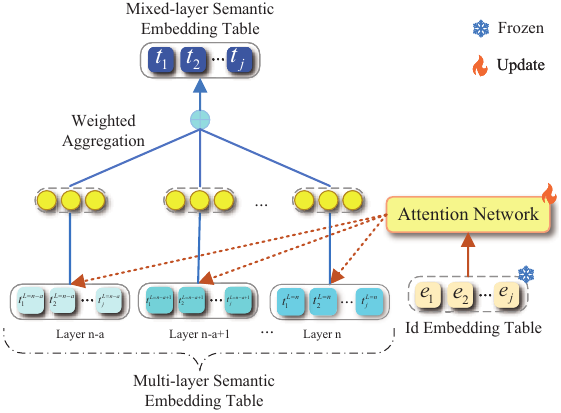}
    \caption{The structure of the fusion block with Id embeddings as supervision signals.}   \label{fig:fusion_block}
\end{figure}
\subsubsection{Fusion Block}\label{sec:fusion_block}
Since each layer of semantic embedding captures distinct aspects of semantic knowledge, we employ a dynamic fusion strategy to integrate the multi-layer semantic embeddings (as illustrated in Fig~\ref{fig:fusion_block}). Within the recommendation model, the ID embedding primarily encodes implicit collaborative signals between users and items~\cite{cheng2024empowering}. Building on this concept, we leverage the ID embedding to guide the weight distribution across different layers of the semantic embeddings, enhancing the model's ability to capture the relationship between user interaction features and semantic information. By doing this, we can better extract the semantically valuable information from each layer by the fusion block for further encoding.
For each item $i$, we concatenate its text embedding from each layer with the corresponding ID embedding $\bm{e}_i \in \bm{E}$, where $\bm{E}$ denotes the ID embedding table. Then we pass this concatenated vector through a \ac{MLP} to obtain the weights corresponding to each layer of the text embedding, which is shown as follows:
\begin{gather}
    w_{i, j}^* = \bm{W}_2 \cdot \text{LReLU}(\bm{W}_1 \cdot [\bm{e}_i \oplus \bm{t}_i^{L=j}] + \bm{b}_1) + \bm{b}_2, \\
    \label{eq:layer_weight}
    w_{i, j} = \frac{\text{exp}(w_{i,j}^*)}{\sum\limits_{j=n-a}^{n}\text{exp}(w_{i,j}^*)},
\end{gather}
where $\bm{W}_1 \in \mathbb{R}^{2{d_V} \times d_V}$, $\bm{W}_2 \in \mathbb{R}^{{d_V} \times 1}$, $\bm{b}_1 \in \mathbb{R}^{d_V}$ and $\bm{b}_2 \in \mathbb{R}^1$ are learnable parameters in the \ac{MLP}, $\text{LReLU} (\cdot)$ denotes the LeakyReLU activation function. It's important to emphasize that, to prevent this process from interfering with the collaborative signals contained in the ID embeddings, we keep the ID embeddings frozen during training in this module. Next, we use the obtained weights to perform a weighted mixing of each layer's semantic embeddings:
\begin{equation}
    \bm{t}_i =  \sum\limits_{j=n-a}^{n} w_{i, j} \cdot \bm{t}_i^{L=j}.
\end{equation}
Then, the mixed-layer semantic embedding table of each domain $\bm{T} = \{\bm{t}_1, \bm{t}_2, \ldots, \bm{t}_M\}$ can eventually be reached. Whenever an epoch of local training is performed, we have to fuse the multi-layer semantic encodings with the weights obtained in Eq.~(\ref{eq:layer_weight}), which is denoted as:
\begin{gather}\label{eq:mix_encoding}
    \bm{X}_F^A = \{\bm{x}_1^A, \bm{x}_2^A, \ldots, \bm{x}_i^A, \ldots, \bm{x}_{M_A}^A\},\\
    \bm{x}_i^A = \sum\limits_{j=n-a}^{n} w_{i, j} \cdot \bm{x}_i^{A, {L=j}}.
\end{gather}
The mixed-layer semantic encodings $\bm{X}_F^A$ will be transmitted to the server for further semantic fusion across domains.

\subsection{Cross-domain Semantic Fusion}\label{sec:semantic_fusion}
To align non-overlapping domains in the semantic space, we combine mixed-layer semantic encodings on the server side by grouping similar semantics into the same cluster. Using an improved version of the K-means algorithm, our objective is to maximize the identification of common semantics across domains. This clustering approach strengthens the shared semantic features through aggregation, thus providing a bridge to connect disjoint domains (as shown in Figure.~\ref{fig:FFMSR} (2)).

\subsubsection{Semantic Clustering}
Compared to the ordinary K-means algorithm~\cite{jin2011k}, our improvements mainly focus on two aspects. First, we adopt the K-means++~\cite{arthur2007k} initialization method to enhance the quality and convergence speed of clustering. Second, we employ weighted aggregation to update the centroids, where the contribution of each point to the centroid is determined by its distance, allowing each point to more accurately find its cluster.

Specifically, upon receiving the mixed-layer semantic encodings from various clients, the server randomly selects centroids from all the encodings based on a predefined number of centroids $K$, using a process similar to the K-means++ initialization method. For each point, the distance to its respective centroid is computed, and this distance determines the probability of the point being selected as the next centroid—the greater the distance, the higher the probability. In the subsequent centroid updating process, rather than using the traditional K-means algorithm's average aggregation approach, we designed a weighted aggregation method based on the distance between each point and the centroid of its respective cluster to update the centroids, and new centroids are generated by aggregation points according to these weights. This approach allows each cluster to concentrate better on semantically related features.
The detailed clustering method is shown in Algorithm~\ref{alg:clustering}.

\begin{algorithm}
\footnotesize
\SetAlgoLined
\LinesNumbered
\caption{$\proc{Clustering Algorithm for Federated Learning}.$}
\label{alg:clustering}
\KwIn{Data from $N$ clients $\{\bm{X}_F^1, \bm{X}_F^2, \ldots, \bm{X}_F^N\}$, number of clusters $K$, max iterations $T$}
\KwOut{Cluster centers $\bm{C} = \{\bm{c}_1, \bm{c}_2, \ldots, \bm{c}_K\}$}

\textbf{Initialization}:\\
Select the first cluster center $c_1$ randomly from the data points.\\
Concatenate the mixed-layer semantic encodings from each client to obtain $\bm{X}_F$.\\

\For{$k = 1$ \KwTo $K$}{
    \ForEach{point $x_i$ in $\bm{X}_F$}{
        compute $d(\bm{x}, \bm{C})$, the distance between $x$ and the nearest center already chosen.\\
        Select a new data point as a center randomly, with probability proportional to $d(\bm{x}, \bm{C})^2$.\\
    }
}

\While{not converge}{
    \textbf{Cluster Assignment}:\\
    \For{each point $\bm{x}_i$ in $\bm{X}_F$}{
        Assign $\bm{x}_i$ to the closest cluster by calculating:
        \[ \text{label}(\bm{x}_i) = \arg\min_{j} \|\bm{x}_i - \bm{c}_j\|^2 \]
    }
    \textbf{Center Update}:\\
    \For{each cluster $j = 1$ \KwTo $K$}{
        Calculate the new center $c_j$ using the weighted average:
        \[ \bm{c}_j = Mean({\sum_{\bm{x}_i \in \bm{C}_j} w_i \bm{x}_i}), \]
        where $w_i = \frac{1}{\|\bm{x}_i - \bm{c}_j\| + \epsilon}$ and $\epsilon$ is a small constant used to prevent the denominator from being 0.\\
    }
}
\end{algorithm}

\subsubsection{Synchronization}\label{sec:synchronization}
Through semantic clustering, we can group similar semantic encodings into the same cluster, and represent the items with it by the corresponding centroid nodes.
That is, we replace the original mixed-layer semantic encoding of each item with the centroid vector of the cluster that it belongs to (we refer to it as the cluster encoding). The cluster encoding will return to the clients as the global domain information, which is denoted by $\bm{X}_C^A$ and $\bm{X}_C^B$ in domains A and B, respectively.

Taking domain A as an example, for the received clustered encoding $\bm{X}_C^A$, we first utilize a \ac{MoE} adapter with the same structure as described in Section~\ref{sec:MoE} to convert it into an embedding table, denoted $\bm{F}^A$.
Then, this global information will collaborate with the local domain data from ID-modality and mixed semantic data for further sequence modeling.

\subsection{Semantic Filtering}\label{sec:semantic_filter}
The clustered encoding received from the server includes a greater amount of common semantic information from various domains, integrating insights to create a comprehensive view of item characteristics. However, using this encoding directly in the local model might introduce noise and degrade performance. To ensure that the local client benefits from the most relevant knowledge, we need to apply local adaptation using semantic filters (as shown in Figure.~\ref{fig:FFMSR} (3)).

\subsubsection{Filter Layer}\label{sec:filter_layer}
Taking user ${u_i^A} \in \mathcal{U}^A$ in client A as an example, whose interaction sequence is denoted as $\mathcal{S}_i^A = \{ A_1, A_2,\ldots, A_j,\ldots\, A_{m_i}\}$. 
Three types of representations, i.e., the clustered global embedding from the server, the encoding from the ID-modality, and the mixed-layer semantic embedding, of the items within $\mathcal{S}_i^A$ are achieved.
To obtain item's global semantic information, the lookup operation is applied to the clustered embedding table $\bm{F}^A$. The resulting sequence embedding is denoted as $\bm{F}_i^A = \{\bm{f}_1^A, \bm{f}_2^A, \ldots, \bm{f}_{m_i}^A\}$.
To filter out the irrelevant semantics brought by the federated learning, we devise a \ac{FFT}-based filter layer with a learnable filter parameter in the frequency domain, which allows us to discover more targeted handling of noise in different frequency components of interaction features~\cite{zhou2022filter}.

To be more specific, we first use \ac{FFT} $\mathcal{F}(\cdot)$ to transform the resulting embedding of the corresponding items into the frequency domain space $\mathbb{C}$:
\begin{equation}\label{eq:FFT}
    {\bm{Y}}_i^A = \mathcal{F}(\bm{F}_i^A) \in \mathbb{C}^{{m_i} \times {d_V}},
\end{equation}
where ${\bm{Y}}_i^A$ denotes the spectrum of $\bm{F}_i^A$. Then we perform the filtering operation by a learnable filtering parameter $\bm{W}_c$ for each dimension of $\bm{Y}_i^A$, which is denoted as:
\begin{equation}
    \widetilde{\bm{Y}}_i^A = \bm{W}_c \odot {\bm{Y}}_i^A, 
\end{equation}
where $\odot$ is the element-wise multiplication. Finally, we transform the filtered spectrum $\widetilde{\bm{Y}}_i^A$ into the domain space by finding the inverse of \ac{FFT}, denoted $\mathcal{F}^{-1}(\cdot)$:
\begin{equation}\label{eq:IFFT}
    \widetilde{\bm{F}}_i^A \leftarrow \mathcal{F}^{-1}(\widetilde{\bm{Y}}_i^A) \in \mathbb{R}^{{m_i} \times {d_V}},
\end{equation}
where $\widetilde{\bm{F}}_i^A = \{\tilde{\bm{f}}_1^A, \tilde{\bm{f}}_2^A, \ldots, \tilde{\bm{f}}_{m_i}^A\}$ is the filtered clustered embedding vector.

\subsubsection{Gating Mechanism}
To obtain user ${u_i^A}$'s representations in ID-modality, another lookup operation is conducted on $\bm{E}^A$.
The resulting item embeddings are denoted as $\bm{E}_i^A = \{\bm{e}_1^A, \bm{e}_2^A, \ldots, \bm{e}_{m_i}^A\}$.
However, as the ID-modality often contains redundant collaborative signals,  we further design a gating mechanism to dynamically re-allocate the contribution of the ID-modality.

For a given item $\bm{e}_j^A \in \bm{E}_i^A$, we denote the gating operation performed on it as: 
\begin{gather}\label{eq:id_gate1}
    {\bm{e}_j^A}^{\prime} = \bm{w}_j^{\text{ID}} \cdot \bm{e}_j^A, \\
    \label{eq:id_gate2}
    \bm{w}_j^{\text{ID}} = \text{Sigmoid}(\bm{e}_j^A \cdot \bm{W}_j^{\text{ID}}),
\end{gather}
where $\bm{W}_j^{\text{ID}}: \mathbb{R}^{d_V} \to \mathbb{R}$ is the linear layer used to determine the weights of ID embedding. We denote the output of the gating mechanism for the given ID embedding by ${\bm{E}_i^A}^{\prime} = \{{\bm{e}_1^A}^{\prime}, {\bm{e}_2^A}^{\prime}, \ldots, {\bm{e}_{m_i}^A}^{\prime}\}$.

\begin{figure}[t]
    \centering    \includegraphics[width=6cm]{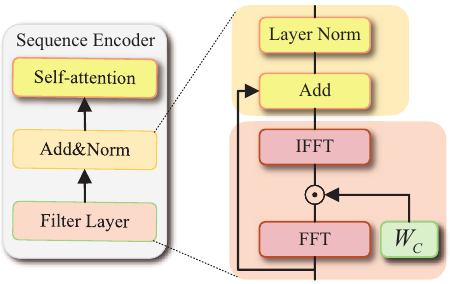}
    \caption{The improved sequence encoder with \ac{FFT}-based filter layer.}    \label{fig:sequence_encoder}
\end{figure}

\subsubsection{Sequence Encoder}\label{sec:sequence_encoder}
After obtaining the clustered embedding $\widetilde{\bm{F}}_i^A = \{\tilde{\bm{f}}_1^A, \tilde{\bm{f}}_2^A, \ldots, \tilde{\bm{f}}_{m_i}^A\}$, the ID embedding ${\bm{E}_i^A}^{\prime} = \{{\bm{e}_1^A}^{\prime}, {\bm{e}_2^A}^{\prime}, \ldots, {\bm{e}_{m_i}^A\}}^{\prime}$, and the mixed-layer semantic embedding $\bm{T}_i^A = \{\bm{t}_1^A, \bm{t}_2^A, \ldots, \bm{t}_{m_i}^A\}$, 
we fuse them by directly adding them together.
The combined item embedding vector is denoted as $\bm{V}_i^A = \{\bm{v}_1^A, \bm{v}_2^A, \ldots, \bm{v}_j^A, \ldots, \bm{v}_{m_i}^A\}$.
For any $\bm{v}_j^A \in \bm{V}_i^A$, we reach its representation as follows:
\begin{equation}\label{eq:combine}
    \bm{v}_j^A = \tilde{\bm{f}}_j^A + {\bm{e}_j^A}^{\prime} + \bm{t}_j^A.
\end{equation}
By doing this, the fused embedding vector can have features from multiple sources, i.e., the clustered embedding, Id modality, and text modality, which contain more information than only using the features in ID-modality. 
But this also means it may introduce irrelevant features than a single modality.
To deal with this, 
we then improve the Transformer-style sequence encoder~\cite{kang2018self}
by introducing a Filter Layer (similar to in Eqs.~(\ref{eq:FFT})-(\ref{eq:IFFT})) to conduct noise filtering (as illustrated in Fig~\ref{fig:sequence_encoder}).

Specifically, we adopt the same structure as in Eqs.~(\ref{eq:FFT})-(\ref{eq:IFFT}) in the filter layer of the sequence encoder, and indicate the filtered vector as $\widetilde{\bm{V}}_i^A = \{\tilde{\bm{v}}_1^A, \tilde{\bm{v}}_2^A, \ldots, \tilde{\bm{v}}_{m_i}^A\}$. To prevent training instability and better integrate with the Transformer structure, we use residual connection and layer normalization as in \cite{ba2016layer} after the filter layer~\cite{he2016deep}:
\begin{equation}
    \widetilde{\bm{V}}_i^A = \text{LayerNorm}(\bm{V}_i^A + \text{Dropout}(\widetilde{\bm{V}}_i^A)).
\end{equation}
In our method, the multiple layers are employed for the filter, and $\widetilde{\bm{V}}_i^A$ serves as the input to the next layer of the filter.

Thereafter, we feed the filtered embeddings into a widely used Transformer structure~\cite{kang2018self} to capture users' temporal interests, which mainly consists of a multi-head self-attention layer (denoted as $\text{MHAtt}(\cdot)$) and a position-aware feed-forward neural network (denoted as $\text{FFN}(\cdot)$). More concretely, for $\tilde{\bm{v}}_j^A \in \widetilde{\bm{V}}_i^A$, we first add it to its corresponding positional embedding:
\begin{equation}
    \bm{h}_{j, 0}^A = \tilde{\bm{v}}_j^A + \bm{p}_j.
\end{equation}
Then, we input the vectors at each position that performed this operation feed to MHAtt ($\cdot$) and FFN ($\cdot$) for encoding:
\begin{gather}
    \bm{H}^{A,l} = [\bm{h}_0^{A, l}; \ldots; \bm{h}_n^{A, l}],\\
    \bm{H}^{A, {l+1}} = \text{FFN}(\text{MHAtt}(\bm{H}^{A, l}), l\in\{1,2, \ldots, L\},
\end{gather}
where $\bm{H}^{A,l}$ denotes the output representation of the $l$-th layer, $L$ is the total layer number of Transformer. We take the hidden state $\bm{h}_i^A = \bm{h}_n^L$ at the $n$-th position as the final sequence representation.

\subsection{Learning Objective}
The training process of \ac{FFMSR} can be divided into two phases: the federated pre-training phase and the local client fine-tuning phase after federated learning. The specific details can be found in Algorithm~\ref{alg:FFMSR}.
\subsubsection{Pre-training in Local Clients}\label{sec:local_training}
Given the sequence representation $\bm{h}_i^A$ (take domain A as an example), we represent the prediction probability of the next potential interaction item as:
\begin{gather}
    P({A_{{m_i}+1}}|{\{ A_1, A_2,\ldots\, A_{m_i}\}}) = \text{Softmax}(\bm{h}_i^A \cdot {\bm{I}^A}^{\top}),\\
    \bm{I}^A = \bm{T}^A + {\bm{E}^A},
\end{gather}
where ${\bm{E}^A}$ represents the ID embedding in domain A. Then we use the cross-entropy as our initial training loss in local clients:
\begin{equation}\label{eq:pred_loss}
    \mathcal{L}_{CE}^A =  - \frac{1}{{\left| {{\mathcal{S}_A}} \right|}}\sum\limits_{{\mathcal{S}_i^A} \in {\mathcal{S}_A}} {\log P({A_{i + 1}}|{\mathcal{S}_i^A}}),
\end{equation}
where $\mathcal{S}_A$ is the training set in domain A.

Moreover, given that semantic embeddings and ID embeddings in the model are used to capture different features of items, we apply an enforced orthogonality loss to these two modalities to enhance the diversity and specificity between different feature aspects:
\begin{equation}\label{eq:or_loss1}
    \mathcal{L}_{orthogonal}^A = \left\|{\bm{t}_1^{A}}^{\top} \bm{e}_1^A\right\|_{2} + \left\|{\bm{t}_1^A}^{\top} \bm{e}_1^A\right\|_{2} + \cdots + \left\|{\bm{t}_{M_A}^A}^{\top} \bm{e}_{M_A}^A\right\|_{2},
\end{equation}
where $||\cdot||_{2}$ is the Euclidean norm of a vector. 

Then, the updated training loss in the local client for domain A can be reached as:
\begin{equation}
    \mathcal{L}^A = \mathcal{L}_{CE}^A + \mathcal{L}_{orthogonal}^A.
\end{equation}

\subsubsection{Fine-tuning in Local Clients after Federated Learning}\label{sec:local_finetuning}
Upon finishing the federated training phase, we ensure the thorough training of local models by keeping the cluster embeddings, learned during federated training, fixed while allowing other components to be updated. We persist with training until the model converges. To mitigate the risk of excessive loss of global knowledge embedded in the clustered codes due to local training, we incorporate an orthogonal loss between these codes and ID embeddings. 

This approach further amplifies the complementary and distinctive qualities of the various types of embeddings:
\begin{equation}
    {\mathcal{L}_{orthogonal}^A}^{\prime} = \left\|{\bm{f}_1^{A}}^{\top} \bm{e}_1^A\right\|_{2} + \left\|{\bm{f}_1^A}^{\top} \bm{e}_1^A\right\|_{2} + \cdots + \left\|{\bm{f}_{M_A}^A}^{\top} \bm{e}_{M_A}^A\right\|_{2}.
\end{equation}

The fine-tuning loss of local clients for domain A can be denoted as:
\begin{equation}\label{eq:finetuning_loss}
    \mathcal{L}^A = \mathcal{L}_{CE}^A + \mathcal{L}_{orthogonal}^A + {\mathcal{L}_{orthogonal}^A}^{\prime}.
\end{equation}

\begin{algorithm}[h]
\footnotesize
\SetAlgoLined
\caption{$\proc{The Training Process of FFMSR}$}
\label{alg:FFMSR}
\LinesNumbered
\KwIn{Interaction sequence from two clients, $\mathcal{S}^A$ and $\mathcal{S}^B$; descriptions of all items, $\mathcal{T}$.}
\KwOut{Next-item predictions for each user in the client.}
\textbf{Stage 1: Federated Learning:} \\
\textbf{Initialization:} Obtain the multi-layer semantic encoding $\bm{x}^L$ for each item using the method described in \ref{sec:multi_layer_encoding}.\;
\For{each epoch $i$ with $i = 1, 2, \ldots$}{
\textbf{Client\_Executes:} \\
\For{each client j}{
    Obtain the multi-layer semantic embedding table via \ac{MoE} adapter as described in Sec.~\ref{sec:MoE} \;
    Obtain the mixed-layer semantic embedding table $\bm{T}^j$ via fusion block as described in Sec.~\ref{sec:fusion_block}. \\
     Receive the clustered encoding $\bm{X}_C^j$ from the server \;
    Get the combined vector via Eq.~(\ref{eq:combine}). \\
    Obtain the user representation vectors $\bm{h}^j$ via the method described in Sec.~\ref{sec:sequence_encoder} \;
    Local training by computing prediction loss $\mathcal{L}_{CE}^j$ of Eq.~(\ref{eq:pred_loss}) and orthogonal loss $\mathcal{L}_{orthogonal}^j$ of Eq.~(\ref{eq:or_loss1}) \;
    Transmit the mixed-layer semantic encodings $\bm{X}_F^j$ obtained with Eq.~(\ref{eq:mix_encoding}). \\
}
\textbf{Server\_Executes:} \\
Clustering the semantic encodings transmitted by each client by Algorithm~\ref{alg:clustering} \;
Distribute the clustered encodings back to individual clients as described in Sec.~\ref{sec:synchronization}. \\
}
\textbf{Stage 2: Local Fine-tuning:} \\
\While{not converge}{
Freezing the clustering encoding and training with the loss in Eq.~(\ref{eq:finetuning_loss}). \\
}
\end{algorithm}

%% file: 04-setup.tex
\section{Experimental Setup}
In this section, we outline the key questions addressed in the experiments and offer detailed information on the dataset, evaluation protocols, baselines, and the implementation of \ac{FFMSR}.

\subsection{Research Questions}
We thoroughly assess our model by examining the following issues:
\begin{itemize}
    \item[\textbf{RQ1}] How does our method compare to various state-of-the-art recommendation methods?
    \item[\textbf{RQ2}] What effect do the core components of \ac{FFMSR}—namely, clustered encoding, the filter layer, orthogonal loss, and the gating mechanism—have on the performance of recommendations?
    \item[\textbf{RQ3}] How do varying hyper-parameters in \ac{FFMSR} influence performance?
    \item[\textbf{RQ4}] What is the effect on \ac{FFMSR} when different clustering algorithms are used on the server?
\end{itemize}

\begin{table}[h] %
	\caption{The statistics of our datasets, where ``Avg.~$n$'' denotes the average length of the user interaction sequence.
	}
	\label{tab:dataset}
    \small
    \setlength{\tabcolsep}{4pt}
	\begin{tabular}{l *{5}{r}}
		\toprule
		\textbf{Datasets} & \textbf{\#Users} & \textbf{\#Items} & \textbf{\#Inters.} & \textbf{Avg.~$n$} \\
		\midrule
  		\textbf{Office}      & 87,436 & 25,986 & 684,837 & 7.84  \\
            \textbf{Arts}        & 45,486 & 21,019 & 395,150 & 8.69  \\
            \midrule
            \textbf{OnlineRetail} & 16,520 & 3,469 & 519,906 & 26.90  \\
		\textbf{Pantry}      & 13,101 &  4,898 & 126,962 & 9.69  \\
		\bottomrule
	\end{tabular}
\end{table}
\subsection{Datasets}
We conduct experiments on two pairs of domains in different datasets, i.e., ``\textit{Office-Arts}" and ``\textit{OnlineRetail-Pantry}", to evaluate our model. Office, Arts, and Pantry are all from Amazon\footnote{https://nijianmo.github.io/amazon/index.html}, which is a widely used real-world dataset in recommender systems, which contains user reviews and metadata of products in the Amazon platform.
The OnlineRetail dataset is from an online retail company in the UK\footnote{https://www.kaggle.com/carrie1/ecommerce-data}, containing user purchase records and brief descriptions of products.

To assess the cross-domain performance of FFMSR, we take Office and Arts in the Amazon as one pair of domains, achieving our first "Office-Arts" dataset, and OnlineRetail and Pantry from different platforms into another. Given our focus on sequential recommendation tasks, we arrange user interaction sequences in chronological order. To maintain the non-overlapping characteristic, we do not apply any alignment processing to users. Concretely, we do not match users between domains by their identification information. We reassigned IDs to users and made them all start numbering from 0.
For item text in the Amazon dataset, we adopt standard practices from previous methods by using titles, categories, and brands as item descriptions. In the OnlineRetail dataset, we utilize the default descriptions provided as item text.
To address the issue of overly sparse data, we implement a five-core strategy that filters out users and items with fewer than five interactions.
The statistics of the resulting datasets are shown in Table~\ref{tab:dataset}.

\subsection{Evaluation Protocols}
To assess our method for next-item prediction on sequential interaction data, we use a leave-one-out strategy to divide user interaction sequences. Specifically, the last three items in each sequence are allocated for training, validation, and testing. During training, the model uses all interactions except the target item. For validation and testing, we rank all items based on their ground-truth item using full sorting, and average the results across all users to obtain the final outcomes. 
We utilize Recall@N and NDCG@N as our evaluation metrics, where N is set as 10 and 50 in experiments.

\subsection{Baselines\label{baselines}}
We compare \ac{FFMSR} with the following three types of baselines:

ID-only recommendation methods:
\begin{itemize}
    \item \textbf{SASRec}~\cite{kang2018self}: This is a self-attention-based method that employs the Transformer architecture to encode the user’s historical interactions and generate personalized item embeddings.
    \item \textbf{Bert4Rec}~\cite{sun2019bert4rec}: This is a bidirectional self-attention-based model that adapts the BERT framework to sequential recommendation. It uses a cloze task to mask and predict items in the sequence by joint conditioning on both left and right contexts.
    \item \textbf{FEARec}~\cite{du2023frequency}: This method proposes a novel frequency-enhanced hybrid attention network. It introduces an auto-correlation-based attention mechanism for periodic features and combines time and frequency domain attention. 
\end{itemize}

Cross-domain recommendation methods:
\begin{itemize}
    \item \textbf{CCDR}~\cite{xie2022contrastive}: This method devises an intra-domain contrastive learning (intra-CL) task and three inter-domain contrastive learning (inter-CL) tasks for cross-domain matching. One of the contrastive loss terms is used to align overlapping users between domains, and due to the non-overlapping user setup in our experiments, we removed this loss when conducting experiments with this method.
    \item \textbf{RecGURU}~\cite{li2022recguru}: This is an adversarial learning model that leverages generalized user representations (GURs) for cross-domain recommendation. Due to the scenario we set, we removed the constraint of overlapping users.
\end{itemize}

ID-Text recommendation methods:
\begin{itemize}
    \item \textbf{FDSA}~\cite{zhang2019feature}: This method uses two different transformer encoders to get the text representation and the ID representation separately and eventually fuses them together with a linear layer.
    \item \textbf{S$^3$Rec}~\cite{zhou2020s3}:  This is a self-supervised learning model that utilizes the mutual information maximization principle to learn the correlations among attributes, items, subsequences, and sequences.  
    \item \textbf{UniSRec}~\cite{hou2022towards}: This method utilizes textual information to get a generic textual representation of items. For a fair comparison with our approach, we trained UniSRec from scratch and used a version that uses both ID embedding and text embedding.
\end{itemize}

Federated cross-domain recommendation method:
\begin{itemize}
    \item \textbf{PFCR}~\cite{guo2024prompt}: This approach is designed for the same scenario with \ac{FFMSR}, which utilizes vector quantification to create a shared embedding between domains and transfer cross-domain information through this embedding. We compare \ac{FFMSR} with it to the version that employs the full prompting strategy.
    \item  \textbf{FedDCSR}~\cite{zhang2024feddcsr}: This approach aims to protect the data privacy of the domain, by introducing an inter-intra domain sequence representation disentanglement, which decomposes user sequence features into domain-shared and domain-exclusive types and learns global features across different domains through federated learning.
\end{itemize}

\subsection{Implementation Details}
We implement \ac{FFMSR} on the basis of the PyTorch framework, and all experiments are conducted on the NVIDIA RTX 3090 GPU. In the deep semantic extraction module, the original semantic encoding output by \ac{PLM} has a dimension of 768. We choose to retain the semantic encodings from the last 3 layers, and the hidden size of the \ac{FFMSR} is set to 300. For the cross-domain semantic fusion module, considering the disparity in data volume between the pairs of datasets used in the experiment, the number of centroids for clustering on the server is set to 120 for ``\textit{Office-Arts}" and 90 for ``\textit{OnlineRetail-Pantry}". For the semantic filtering module, the sequence encoder's filtering layers are set to 2, while only one layer is used for filtering the clustering embedding. During the training phase, we employ the Adam optimizer with a learning rate of 0.001 and set the batch size to 1024. For the federated pre-training phase, we specify 5 training rounds. In the local client fine-tuning phase after federated training, we apply an early stopping strategy with a patience of 10. We select the model with the best Recall@10 on the validation set for testing. For all baselines, we set the batch size as 1024, and the hyper-parameters within the models strictly follow their original settings.

%% file: 05-result.tex
\begin{table*}[t]
\centering
\small
\caption{Comparison results on \textit{Office-Arts} and \textit{OnlineRetail-Pantry}.}
\label{tab:perf}
\fontsize{8.5}{12}\selectfont
\scalebox{0.82}{
\setlength{\tabcolsep}{0.85mm}{
\begin{tabular}{llcccccccccccc}
\toprule
\multirow{2}{*}{\textbf{Dataset}} & \multicolumn{1}{c}{\multirow{2}{*}{\textbf{Metric}}}   & \multicolumn{3}{c}{\textbf{ID-Only}}  &\multicolumn{2}{c}{\textbf{Cross-Domain}}  & \multicolumn{3}{c}{\textbf{ID-Text}} & \multicolumn{2}{c}{\textbf{Federated}}  & \multicolumn{1}{c}{\textbf{Ours}} & \multirow{2}{*}{Improv.} \\ \cmidrule(lr){3-5} \cmidrule(lr){6-7} \cmidrule(lr){8-10} \cmidrule(lr){11-12} \cmidrule(lr){13-13}  
 & &\textbf{SASRec} & \textbf{Bert4Rec} & \textbf{FEARec} & \textbf{CCDR} & \textbf{RecGURU}  & \textbf{FDSA}       & \textbf{S$^3$Rec}       & \textbf{UniSRec} &\textbf{FedDCSR} & \textbf{PFCR}  & \textbf{FFMSR} &                                      \\ \midrule
\multirow{4}{*}{Office} & Recall@10 & 0.1157 & 0.0693 & 0.1155 & 0.0549 & 0.1145  & 0.1091  & 0.1030 & {\ul 0.1257} & 0.1022 & 0.1206 & \textbf{0.1312}* & +4.38\%      \\
& NDCG@10 & 0.0663 & 0.0554 & 0.0756 & 0.0290 & 0.0768  & 0.0821 & 0.0653 & {\ul 0.0843} & 0.0787   & 0.0782 & \textbf{0.089}*  & +5.58\%  \\
& Recall@50 & 0.1799  & 0.1182  & 0.1804 & 0.1095 & 0.1757  & 0.1697 & 0.1613 & {\ul 0.1967} & 0.1525 & 0.1938 & \textbf{0.2069}*  & +5.19\%   \\ 
& NDCG@50 & 0.0803  & 0.0617  & 0.0897 & 0.0401 & 0.0901  & 0.0953 & 0.0780 & {\ul0.0998} & 0.0896 & 0.0941 & \textbf{0.1054}* & +5.61\%   \\ \cline{2-14}

\multirow{4}{*}{Arts} & Recall@10 & 0.1048 & 0.0638 & 0.1155 & 0.0671 & 0.1084  & 0.1016 & 0.1003 & {\ul 0.1219} & 0.0852 & 0.1153 & \textbf{0.1257}*  & +3.12\%   \\
& NDCG@10 & 0.0557 & 0.0364 & 0.062 & 0.0348 & 0.0651  & 0.0671 & 0.0601 & {\ul 0.0679} & 0.0591 & 0.0677 & \textbf{0.0711}*  & +4.71\%    \\
& Recall@50 & 0.1948  & 0.1318  & 0.2031 & 0.1478 & 0.1979  & 0.1887 & 0.1888 & {\ul 0.2323} & 0.1505 & 0.2199 & \textbf{0.2377}* & +2.32\%      \\ 
& NDCG@50 & 0.0753  & 0.0510  & 0.0822 & 0.0523 & 0.0845  & 0.0860 & 0.0793 & {\ul 0.0919} & 0.0733 & 0.0904  & \textbf{0.0955}*  & +3.92\%   \\ \midrule

\multirow{4}{*}{OR}& Recall@10 & 0.1484 & 0.1392 & 0.1351 & 0.1347 & 0.1467  & 0.1487 & 0.1418 & {\ul 0.1591} & 0.1407 & 0.1561  & \textbf{0.1655}*  & +4.02\%   \\
& NDCG@10 & 0.0684 & 0.0642 & 0.0659 & 0.0658 & 0.0535  & 0.0715 & 0.0654 & {\ul 0.0752} & 0.0688 & 0.0739  & \textbf{0.0763}* & +1.42\% \\
& Recall@50 & 0.3921  & 0.3668  & 0.3367 & 0.3587 & 0.3885  & 0.3748 & 0.3702 & 0.3946 & 0.3367 & {\ul 0.3982} & \textbf{0.411}* & +3.21\%   \\ 
& NDCG@50 & 0.1216  & 0.1137  & 0.1099 & 0.1108 & 0.1188  & 0.1208 & 0.1154 & 0.1266 & 0.1114 & {\ul 0.1266}  &\textbf{0.13}* & +2.69\%    \\ \cline{2-14}

\multirow{4}{*}{Pantry} & Recall@10 & 0.0467 & 0.0281 & 0.0484 & 0.0408 & 0.0469  & 0.0414 & 0.0444 & {\ul 0.0677} & 0.0292 &0.0616 & \textbf{0.0695}*  &+2.66\%  \\
& NDCG@10 & 0.0207 & 0.0142 & 0.0214 & 0.0203 & 0.0209  & 0.0218 & 0.0214 & {\ul 0.0309} & 0.0141 &0.0293  &\textbf{0.0326}* &+5.50\%   \\
& Recall@50 & 0.1298  & 0.0984  & 0.1298 & 0.1262 & 0.1269  & 0.1198 & 0.1315 & {\ul 0.1787} & 0.0875 & 0.1591  & \textbf{0.1818}*  &+1.73\%  \\ 
& NDCG@50 & 0.0385  & 0.0292  & 0.0389 & 0.0385 & 0.0379  & 0.0385 & 0.0400 & {\ul 0.055} & 0.0265 & 0.0503  & \textbf{0.0568}*  & +3.27\% \\ \bottomrule
\end{tabular}
}}
\begin{tablenotes}  
        \footnotesize       
        \item Significant improvements over the best baseline results are marked with * (t-test, $p$< .05).
\end{tablenotes}
\end{table*}

\section{Experimental Results (RQ1)}
The results of the comparison across all datasets are shown in Table~\ref{tab:perf}, leading to the following conclusions: 1) Our \ac{FFMSR} method consistently outperforms all baselines across every metric and dataset. This indicates that \ac{FFMSR} can effectively capture the crucial information from both ID and text modalities, and leverage semantic information to bridge non-overlapping domains. 2) \ac{FFMSR} surpasses all ID-based methods (e.g., SASRec, Bert4Rec, FEARec) as well as those based on attention mechanisms and \ac{FFT} models (i.e., FEARec). The model’s advantage lies in its use of semantic information to facilitate the transfer of common knowledge across domains. 
\ac{FFMSR} outperforms traditional attention and \ac{FFT} -based methods, indicating the effectiveness of enhancing user feature extraction by integrating \ac{FFT}-based filter layers, which can lead to more accurate results.
3) \ac{FFMSR} outperforms the \ac{CDR} methods (i.e., CCDR, and RecGURU) that rely on overlapping users, demonstrating the advance of \ac{FFMSR} in handling the information transfer in non-overlapping scenarios.
The method that relies solely on textual similarity for transferring semantic knowledge enables us to capture users' common interest at the cluster-level, which facilitates the cross-domain recommendation.
4) \ac{FFMSR} is superior to other ID-Text methods (i.e., FDSA, S$^3$Rec, UniSRec), indicating the usefulness of our federated semantic learning framework in handling the ID and text modalities.
By employing deep semantic extraction, cross-domain semantic fusion, and semantic filter, we can do well in the privacy-preserving non-overlapping cross-domain recommendation scenario.
5) Compared to PFCR, the SOTA method in \ac{PPCDR}, \ac{FFMSR} consistently shows superior results.
This indicates the superiority of learning semantics from the source in \ac{FFMSR}, rather than using the quantizing semantic encodings, which may lead to a loss of rich semantic content.
Moreover, we only engage in federated learning at the semantic encoding level and keep the federated knowledge fixed on the client side, thereby avoiding the impact of heterogeneity.


%% file: 06-analysis.tex
\section{Model Analysis\label{model_analysis}}

In this section, we examine the effect of each module in our model using ablation studies to address RQ2. Subsequently, we examine the impact of key hyper-parameters within the model to answer RQ3. Lastly, we apply various standard clustering techniques to structure our experiments with the goal of answering RQ4.

\begin{table}[h]
    \caption{Ablation studies on \textit{Office-Arts} and \textit{OnlineRetail-Pantry}.}
    \centering
    \scriptsize
    \scalebox{0.85}{
    \setlength{\tabcolsep}{0.8mm}{
    \begin{tabular}{lcccccccc|cccccccc}
    \toprule
    \multicolumn{1}{c}{\multirow{2}[4]{*}{\textbf{Variants}}} & 
    \multicolumn{4}{c}{\textbf{Office }} & 
    \multicolumn{4}{c|}{\textbf{Arts }} &
     \multicolumn{4}{c}{\textbf{OnlineRetail }} &
      \multicolumn{4}{c}{\textbf{Pantry }} \\
    \cmidrule{2-17}
          & \multicolumn{2}{c}{Recall} & \multicolumn{2}{c}{NDCG}
          & \multicolumn{2}{c}{Recall} & \multicolumn{2}{c|}{NDCG}
            & \multicolumn{2}{c}{Recall} & \multicolumn{2}{c}{NDCG}
          & \multicolumn{2}{c}{Recall} & \multicolumn{2}{c}{NDCG}\\
          \cmidrule{2-17}
          & @10      & @50   & @10    & @50   & @10    & @50   & @10    & @50  & @10      & @50   & @10    & @50   & @10    & @50   & @10    & @50\\
    \midrule
     w/o semantic fusion &0.1292 &0.2053 &0.0865 &0.103 &0.1231 &0.2362 &0.0693 &0.094 &0.1644 &0.4091 &0.0757 &0.1295 &0.0679 &0.1756 &0.0321 &0.0553 \\
     w/o adaptation filter &0.1298 &0.2028 &0.0887 &0.1046 &0.124 &0.2338 &\textbf{0.0717} &\textbf{0.0956} &\textbf{0.1657} &\textbf{0.422} &0.0741 &\textbf{0.1304} &0.0682 &0.1792 &\textbf{0.0328} &0.0567\\
     w/o encoder filter  &0.1267 &0.1983 &0.0869 &0.1024 &0.1207 &0.23 &0.0688 &0.0926 &0.162 &0.4005 &0.0755 &0.1274 &0.0689 &0.1749 &0.0319 &0.0547 \\
    w/o orthogonal losses &0.1304 &0.204 &0.0857 &0.1017 &0.1245 &0.2346 &0.0678 &0.0918 &0.1654 &0.4155 &0.0744 &0.1291 &0.0673 &0.1719 &0.0304 &0.053 \\ 
     w/o gating &0.1307 &0.2045 &0.0885 &0.1046 &0.124 &0.2355 &0.0687 &0.093 &0.1646 &0.4085 &0.0747 &0.1282 &0.0694 &0.1782 &0.0314 &0.0552 \\
    \midrule
      \textbf{\ac{FFMSR}} &\textbf{0.1312} &\textbf{0.2069} &\textbf{0.089} &\textbf{0.1054} &\textbf{0.1257} &\textbf{0.2377} &0.0711 &0.0955
      &0.1655 &0.411 &\textbf{0.0763} &0.13
      &\textbf{0.0695} &\textbf{0.1818} &0.0326 &\textbf{0.0568}\\
    \bottomrule
    \end{tabular}}}
  \label{tab:ablation}%
\end{table}

\subsection{Ablation Study (RQ2)}
We analyze five variants of \ac{FFMSR} to assess how different components affect model performance:
\begin{itemize}
    \item w/o semantic fusion: This variant of \ac{FFMSR} removes the federated learning component and all mechanisms used to handle clustered encoding, i.e., the filter layer and orthogonal loss mentioned in Section~\ref{sec:filter_layer} and Section~\ref{sec:local_finetuning}. This modification is intended to demonstrate the significance of federated learning in knowledge transfer.
    \item w/o adaptation filter: This version of FFMSR omits the filter layer used for filtering clustered embeddings (as described in Section ~\ref{sec:local_training}). This is to evaluate the impact of the filter layer in the semantic filtering component.
    \item w/o encoder filter: In this variant, the filter layer within the sequence encoder is removed, and a standard transformer structure is utilized as the sequence encoder instead. This change is made to highlight the contribution of the filter layer in modeling user behaviors.
    \item w/o orthogonal losses: This variant removes the orthogonal losses, relying solely on the prediction loss for training. This adjustment aims to demonstrate the role of orthogonal loss in capturing various aspects of item features.
    \item w/o gating: This variant removes the ID embedding weights, intending to assess the effectiveness of the gating mechanism in eliminating unnecessary noises in ID modality.
\end{itemize}

The experimental results are displayed in Table~\ref{tab:ablation}, from which we can draw the following conclusions: 1) \ac{FFMSR} outperforms \ac{FFMSR} (w/o semantic fusion) on all datasets, demonstrating the effectiveness of the federated learning in bridging non-overlapping domains under the privacy-preserving constraint, which is implemented at the semantic-level.
2) \ac{FFMSR} surpasses \ac{FFMSR} (w/o adaptation filter) in most metrics, indicating the usefulness of filtering out irrelevant semantics, which may bring redundant information and harm accurate modeling.
3) \ac{FFMSR} performs better than \ac{FFMSR} (w/o encoder filter) on all metrics, showing the necessity of the encoder filter in sequence modeling.
This also indicates that a simple transformer structure is insufficient to handle the complex features from multiple modalities.
4) The gap between \ac{FFMSR} and \ac{FFMSR} (w/o orthogonal losses) indicates the effectiveness of applying orthogonal losses to maximize the distinctiveness of the local and global semantics, which can learn different aspects of domain semantics.
5) \ac{FFMSR} has better results than \ac{FFMSR} (w/o gating), denoting the necessity of discarding certain information in ID embeddings, which enables us to highlight the benefits of the gating mechanism in improving model performance.
\begin{table}[t]
    \caption{Impact of FFMSR's core components on other ID-Text methods.}
    \centering
    \scriptsize
    \scalebox{0.85}{
    \setlength{\tabcolsep}{0.8mm}{
    \begin{tabular}{ll|cccccccc|cccccccc}
    \toprule
    \multicolumn{1}{c}{\multirow{3}{*}{\textbf{Methods}}} & 
    \multicolumn{1}{c|}{\multirow{3}{*}{\textbf{Variants}}} & 
    \multicolumn{4}{c}{\textbf{Office }} & 
    \multicolumn{4}{c|}{\textbf{Arts }} &
     \multicolumn{4}{c}{\textbf{OnlineRetail }} &
      \multicolumn{4}{c}{\textbf{Pantry }} \\
    \cmidrule{3-18}
          && \multicolumn{2}{c}{Recall} & \multicolumn{2}{c}{NDCG}
          & \multicolumn{2}{c}{Recall} & \multicolumn{2}{c|}{NDCG}
            & \multicolumn{2}{c}{Recall} & \multicolumn{2}{c}{NDCG}
          & \multicolumn{2}{c}{Recall} & \multicolumn{2}{c}{NDCG}\\
          \cmidrule{3-18}
          && @10      & @50   & @10    & @50   & @10    & @50   & @10    & @50  & @10      & @50   & @10    & @50   & @10    & @50   & @10    & @50\\
    \midrule
     \multirow{2}{*}{FDSA} &Origin &0.1091 &0.1697 &0.0821 &0.0953 &0.1016 &0.1887 &0.0671 &0.0860 &0.1487 &0.3748 &0.0715 &0.1208 &0.0414 &0.1198 &0.0218 &0.0385 \\
     &Enhanced &0.1148 &0.1824 &0.0875 &0.0984 &0.1129 &0.2113 &0.069 &0.0903 &0.1568 &0.3945 &0.0755 &0.1239 &0.0589 &0.1517 &0.0273 &0.0404\\
     \midrule
    \multirow{2}{*}{S3Rec} &Origin  &0.1030 &0.1613 &0.0653 &0.0780 &0.1003 &0.1888 &0.0601 &0.0793 &0.1418 &0.3702 &0.0654 &0.1154 &0.0444 &0.1315 &0.0214 &0.0400 \\
    &Enhanced &0.1122 &0.178 &0.0772 &0.0868 &0.1106 &0.2098 &0.0642 &0.0813 &0.1518 &0.3929 &0.0693 &0.1202 &0.0591 &0.1529 &0.0271 &0.0439 \\ 
    \midrule
    \multirow{2}{*}{UniSRec} &Origin &0.1257 &0.1967 &0.0843 &0.0998 &0.1219 &0.2323 &0.0679 &0.0919 &0.1591 &0.3946 &0.0752 &0.1266 &0.0677 &0.1787 &0.0309 &0.0550 \\
      &Enhanced &0.1304 &0.2035 &0.0882 &0.1043 &0.1231 &0.2347 &0.0697 &0.0941 &0.1638 &0.4063 &0.0763 &0.1286 &0.0687 &0.1804 &0.0321 &0.0564\\
    \bottomrule
    \end{tabular}}}
  \label{tab:enhanced}%
\end{table}
\begin{table}[h]
 \caption{Information loss analysis on \textit{Office-Arts} and \textit{OnlineRetail-Pantry}.}
\centering
\scriptsize
\begin{tabular}{ll|cccc}
\toprule
\multicolumn{1}{c}{\multirow{2}{*}{Dataset}} & \multicolumn{1}{c|}{\multirow{2}{*}{Metric}} & \multicolumn{2}{c}{Recall} &  \multicolumn{2}{c}{NDCG}\\ \cmidrule(lr){3-6}
 & & @10 & @50 & @10 & @50 \\ \midrule
\multirow{2}{*}{Office} & FFMSR-Origin  &0.1292 &0.2053 &0.0865 &0.103 \\
 & FFMSR-PQ & 0.1263 & 0.2015 & 0.0741 & 0.0905 \\ \midrule
\multirow{2}{*}{Arts} & FFMSR-Origin &0.1231 &0.2362 &0.0693 &0.094 \\
 & FFMSR-PQ & 0.119 & 0.2241 & 0.0624 & 0.0853 \\ \midrule
 \multirow{2}{*}{OnlineRetail} & FFMSR-Origin  &0.1644 &0.4091 &0.0757 &0.1295 \\
 & FFMSR-PQ & 0.157 & 0.4087 & 0.0707 & 0.1258 \\ \midrule
 \multirow{2}{*}{Pantry} & FFMSR-Origin  &0.0679 &0.1756 &0.0321 &0.0553 \\
 & FFMSR-PQ & 0.0539 & 0.1479 & 0.0234 & 0.0436 \\ \bottomrule
\end{tabular}
\label{tab:infor_loss}
\end{table}
To further demonstrate the superiority of the ID-Text integration mechanism we propose, we add the two core modules of cross-domain semantic fusion and semantic filtering to other ID-Text methods in the baseline for the experiment. From the results shown in Table~\ref{tab:enhanced}, we can find that after adding the core component, other ID-Text methods have all achieved performance improvements, which proves the generality of the core component in FFMSR.

To analyze the information loss brought by the discrete representations (as the \ac{PQ} method in~\cite{guo2024prompt}), we compare our method FFMSR in local clients (denoted as FFMSR-Origin) with its variant (denoted as FFMSR-PQ) by replacing the item text representation learning method, i.e., \ac{MoE} Adapter and Fusion block, with the \ac{PQ} method in~\cite{guo2024prompt}, and keep all other modules unchanged. The comparison results on all the domains are presented in Table~\ref{tab:infor_loss}. From the results, we can observe that the performance of FFMSR-Origin is significantly better than that of FFMSR-PQ on all the domains, which indicates that our representation learning method can retain more valuable information than the FFMSR-PQ method. This result also demonstrates that discrete vectors can only have limited representation capabilities in obtaining items' text representations.

\begin{figure}[htbp]
    \centering
    \includegraphics[width=11.5cm]{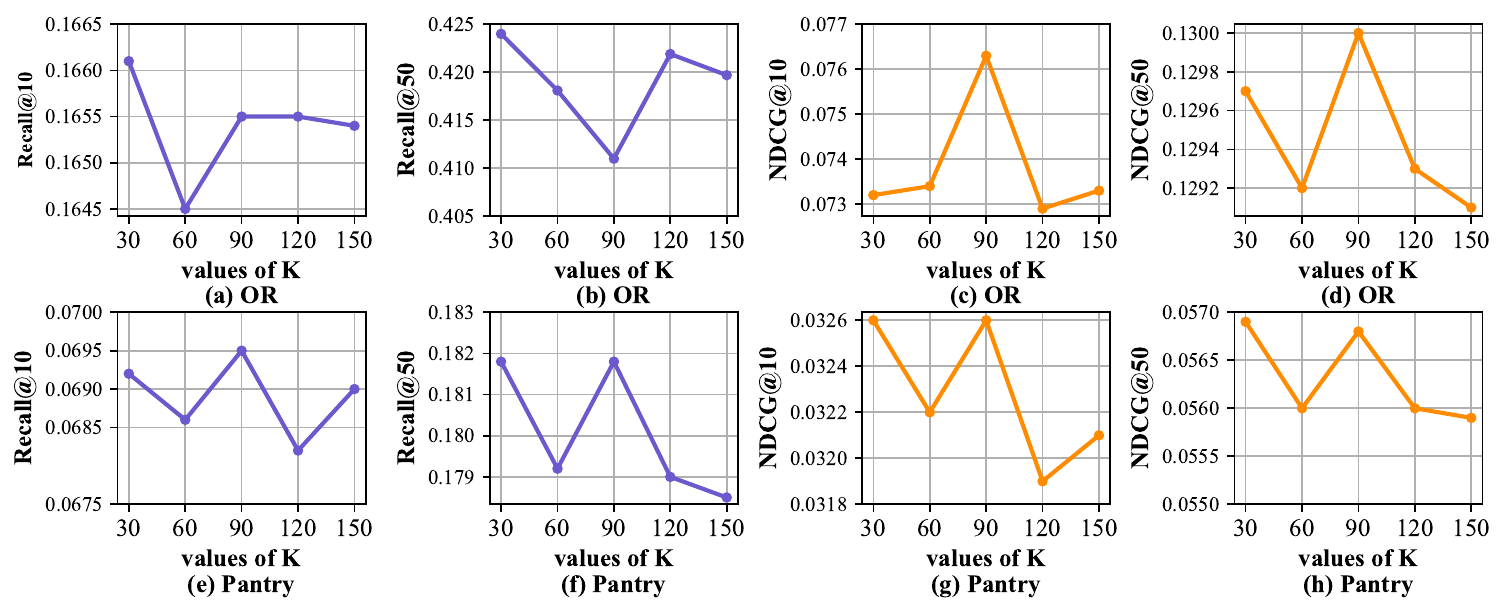}
    \caption{Impact of the number of cluster centers $K$ on \textit{OnlineRetail-Pantry}.}
    \label{fig:num_cluster}
\end{figure}

\begin{figure}[htbp]
    \centering
    \includegraphics[width=11.5cm]{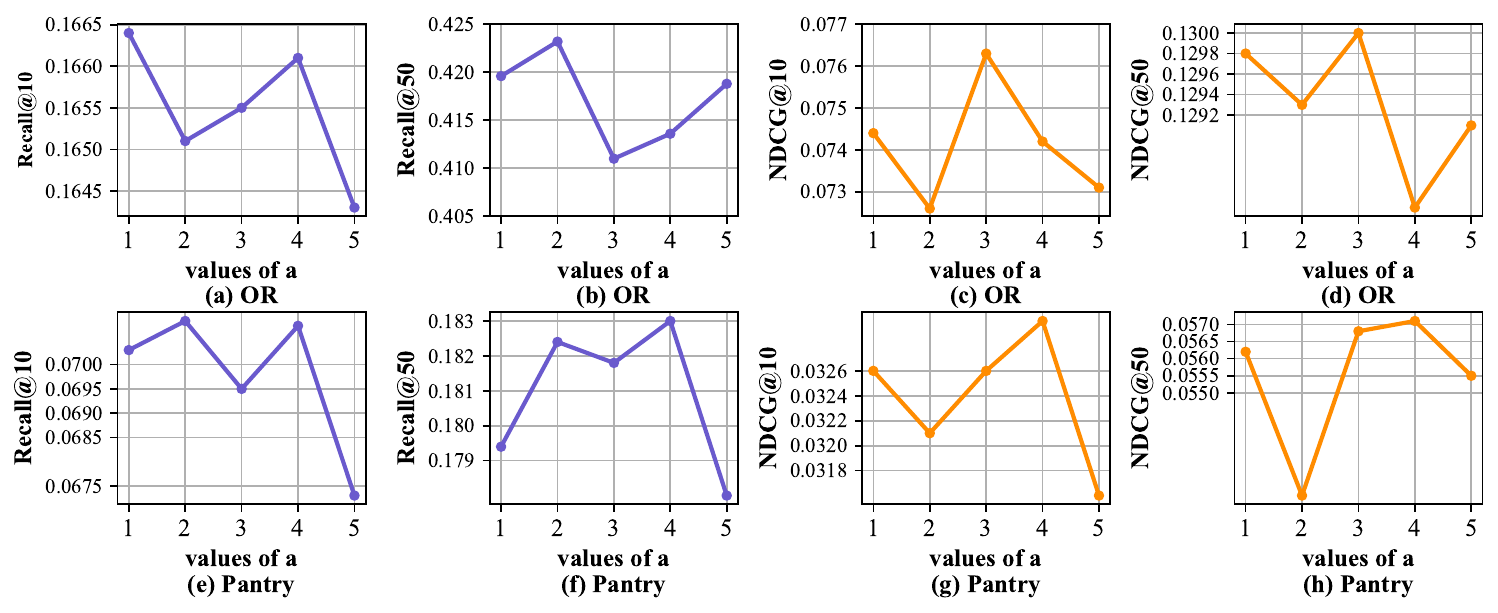}
    \caption{Impact of the number of utilized semantic coding layer in \ac{PLM} $a$ on \textit{OnlineRetail-Pantry}.}
    \label{fig:num_layer}
\end{figure}

\begin{figure}[htbp]
    \centering
    \includegraphics[width=11.5cm]{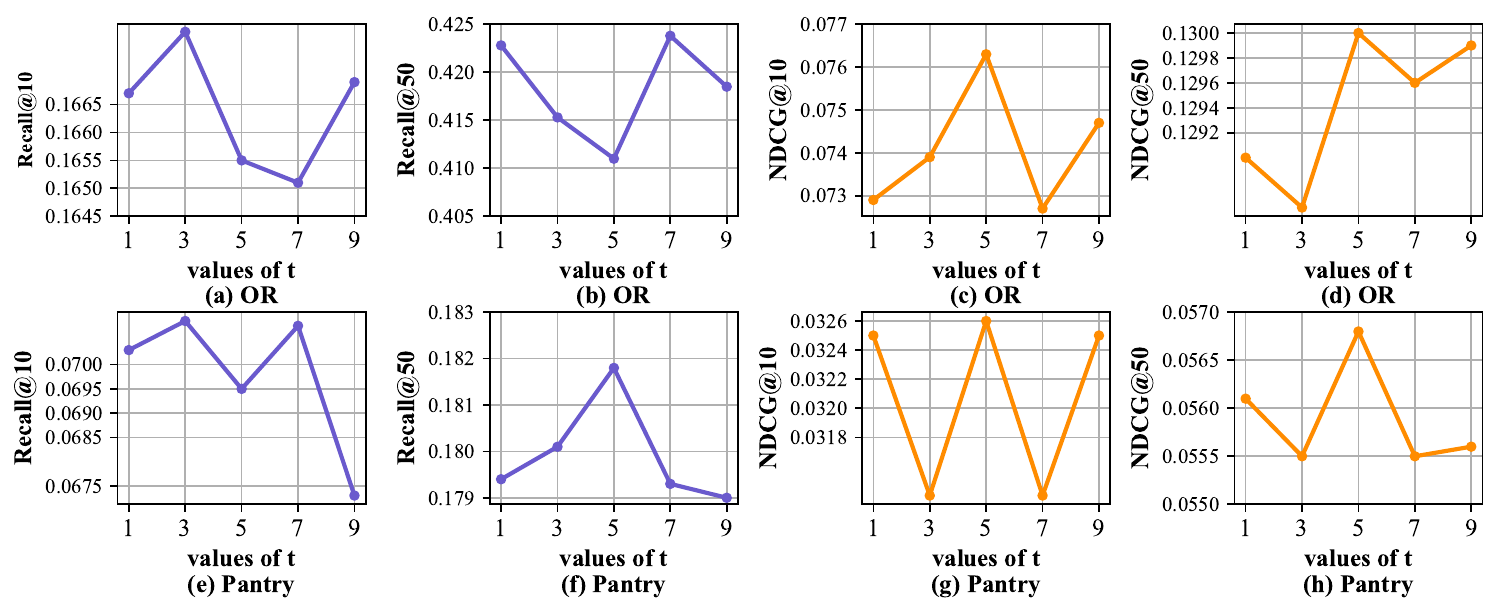}
    \caption{Impact of the number of federated pre-training rounds $t$ on \textit{OnlineRetail-Pantry}.}
    \label{fig:num_epoch}
\end{figure}

\subsection{Hyper-parameter Analysis (RQ3)}\label{sec:hyper_param}
We conduct experiments on three key hyper-parameters, i.e., the number of cluster centroids $K$, semantic encoding layer $a$ used, and rounds of federated pre-training $t$, on \textit{OnlineRetail-Pantry} to investigate their impacts to \ac{FFMSR}. The experimental results are shown in Figs.~\ref{fig:num_cluster}-\ref{fig:num_epoch}, revealing that: 1) 
The optimal performance of \ac{FFMSR} is achieved when $K=30$ or $K=60$ (as illustrated in Fig.~\ref{fig:num_cluster}). 
In other cases, \ac{FFMSR} declines when $K>90$. 
We also find that the best values of $K$ are related to the number of items in the dataset.
Intuitively, too few centroids cannot aggregate all relevant categories, while too many centroids result in insufficient data for each category.
2) The number of semantic encoding layer in \ac{PLM} affects \ac{FFMSR} significantly (as shown in Fig.~\ref{fig:num_layer}).  
From Fig.~\ref{fig:num_layer}, we find that the increase of $a$, it has a positive impact on the NDCG metric. However, the effect on the recall metric varies across different domains. In the OnlineRetail domain, a single layer is optimal due to its sparse textual descriptions, while additional layers may introduce noises.
Conversely, on the Pantry domain, which contains more textual tokens, additional layers can enhance model performance.
But when $a>4$, \ac{FFMSR} performs worse in both domains, suggesting that shallow semantic encoding contains less useful information.
3) An appropriate number of federated rounds significantly influences the effectiveness of clustering encoding (as shown in Fig.~\ref{fig:num_epoch}).
From Fig.~\ref{fig:num_epoch}, we can observe that the best performance occurs when $3<t<5$.

\begin{table}[h]
    \caption{Results of different cluster algorithms on \textit{Office-Arts} and \textit{OnlineRetail-Pantry}.}
    \centering
    \scriptsize
    \scalebox{0.9}{
    \setlength{\tabcolsep}{0.8mm}{
    \begin{tabular}{lcccccccc|cccccccc}
    \toprule
    \multicolumn{1}{c}{\multirow{2}[4]{*}{\textbf{Variants}}} & 
    \multicolumn{4}{c}{\textbf{Office }} & 
    \multicolumn{4}{c|}{\textbf{Arts }} &
     \multicolumn{4}{c}{\textbf{OnlineRetail }} &
      \multicolumn{4}{c}{\textbf{Pantry }} \\
    \cmidrule{2-17}
          & \multicolumn{2}{c}{Recall} & \multicolumn{2}{c}{NDCG}
          & \multicolumn{2}{c}{Recall} & \multicolumn{2}{c|}{NDCG}
            & \multicolumn{2}{c}{Recall} & \multicolumn{2}{c}{NDCG}
          & \multicolumn{2}{c}{Recall} & \multicolumn{2}{c}{NDCG}\\
          \cmidrule{2-17}
          & @10      & @50   & @10    & @50   & @10    & @50   & @10    & @50  & @10      & @50   & @10    & @50   & @10    & @50   & @10    & @50\\
    \midrule
     Agglomerative &0.1308 &0.2058 &0.0886 &0.105 &0.1251 &0.2377 &0.0707 &0.0952 &0.1659 &0.4206 &0.0747 &0.1307 &0.0688 &0.1791 &0.032 &0.0557 \\
     Birch  &0.1309 &0.2068 &0.0887 &0.1053 &0.1255 &0.2385 &0.0706 &0.0952 &0.1657 &0.4203 &0.0737 &0.1296 &0.0684 &0.1811 &0.0321 &0.0564 \\
    K-means &0.1309 &0.2054 &0.0884 &0.1048 &0.124 &0.2372 &0.0708 &0.0955 &\textbf{0.1666} &\textbf{0.4239} &0.0741 &\textbf{0.1305} &\textbf{0.0717} &0.1781 &0.0321 &0.0551 \\ 
     K-means++ &0.1309 &0.2067 &0.0887 &0.1052 &0.1252 &0.2365 &0.0702 &0.0945 &0.1664 &0.4188 &0.074 &0.1294 &0.069 &0.1745 &0.0316 &0.0545 \\
    \midrule
      \textbf{\ac{FFMSR}} &\textbf{0.1312} &\textbf{0.2069} &\textbf{0.089} &\textbf{0.1054} &\textbf{0.1257} &\textbf{0.2377} &\textbf{0.0711} &\textbf{0.0955}
      &0.1655 &0.411 &\textbf{0.0763} &0.13
      &0.0695 &\textbf{0.1818} &\textbf{0.0326} &\textbf{0.0568}\\
    \bottomrule
    \end{tabular}}}
  \label{tab:cluster}%
\end{table}

\subsection{Impact of Different Cluster Algorithms (RQ4)}
We select four prevalent clustering algorithms, i.e., K-means, K-means++, Birch, and Agglomerative, with specified centroid numbers for comparison.
All algorithms are compared with the same centroid number.
For K-means and K-means++, we adopt the same strategy as in \ac{FFMSR}, which conducts clustering until the cluster centers no longer shift. 
For Birch and Agglomerative, as they are not based on iterative optimization, there is no need to set the learning rounds for them.
The experimental results are presented in Table~\ref{tab:cluster}, where we observe that Agglomerative and Birch exhibit similar performance. K-means shows the best performance on most metrics on OnlineRetail.
\ac{FFMSR} achieves the best performance in the majority of metrics on all the datasets, showing the effectiveness of the clustering algorithm in fusing semantics on non-overlapping domains.
\begin{figure}[h]
    \centering
    \includegraphics[width=11.5cm]{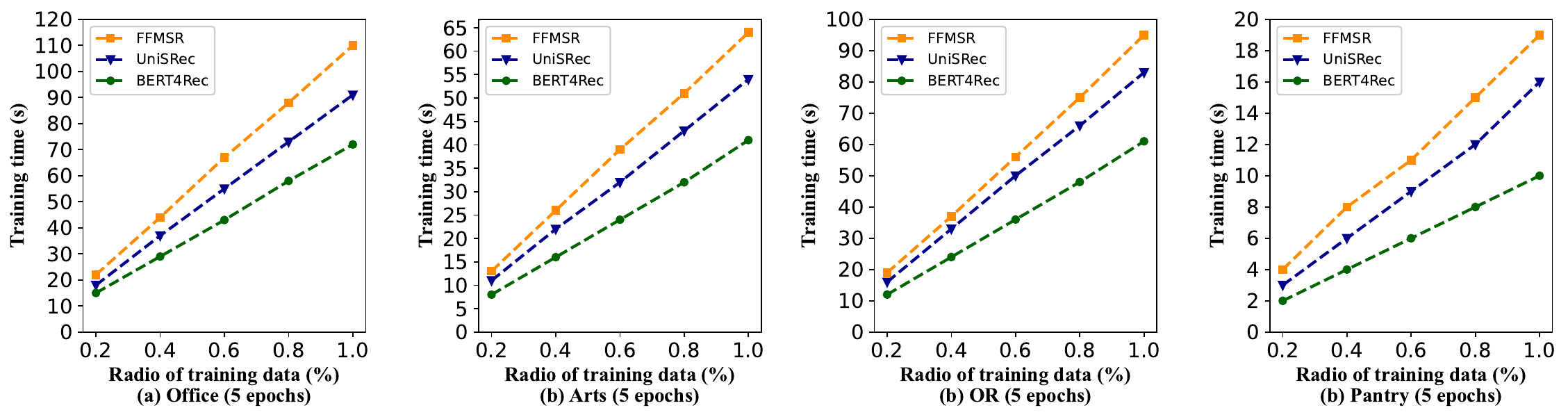}
    \caption{Model training efficiency on \textit{Office-Arts} and \textit{OnlineRetail-Pantry}.}
    \label{fig:time}
\end{figure}
 \begin{figure}[h]
    \centering    \includegraphics[width=11.5cm]{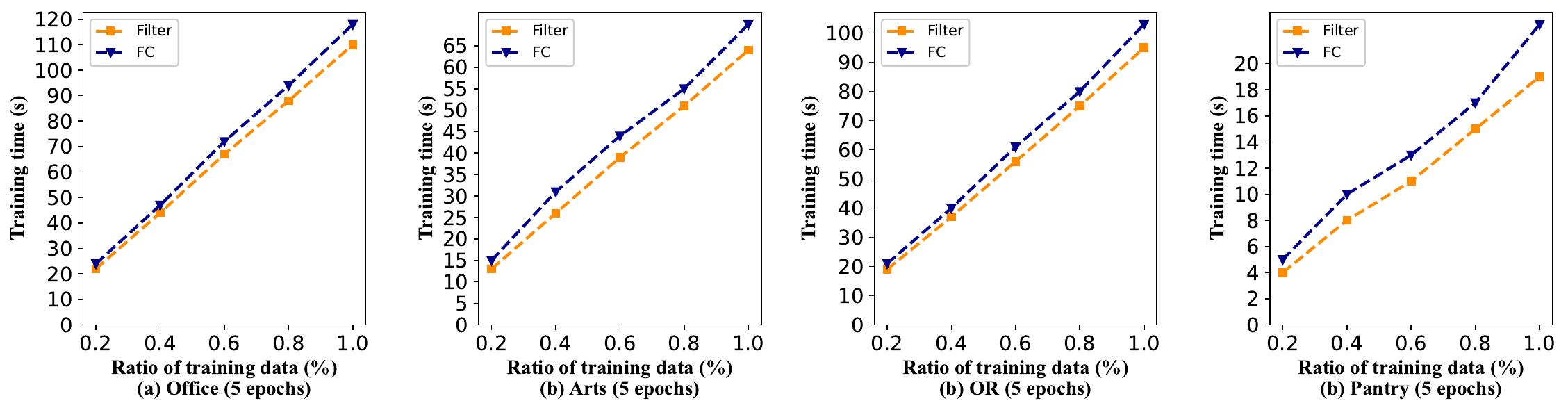}
    \caption{Filter training efficiency on \textit{Office-Arts} and \textit{OnlineRetail-Pantry}.}    \label{fig:filter_time}
\end{figure}
\subsection{Training Efficiency Analysis}
We evaluate the training efficiency of our method by the classic ID-Text method UniSRec~\cite{hou2022towards} and the classic Transformer-style recommendation method Bert4Rec~\cite{sun2019bert4rec} on \textit{Office-Arts} and \textit{OnlineRetail-Pantry}.
The experimental results are presented in Fig~\ref{fig:time}. From this result, we can conclude that: 1) Since \ac{FFMSR} simultaneously uses two modal representations and adds various modules for extracting and utilizing modal knowledge, its training efficiency on each dataset is slightly lower than that of the other two models. But its training time is relatively close to UniSRec, that is, our design does not significantly increase the model complexity.
2) From this result, we also find that the training time of \ac{FFMSR} grows linearly with the increase of the training data ratio, indicating the scalability of our method in applying it to large datasets.

To demonstrate that the filtering method used in \ac{FFMSR} has higher computational efficiency, we compared the computational efficiency of the original FFMSR (referred to as Filter) with a modified version where all the filter layers in \ac{FFMSR} were replaced by fully connected layers of the same number (denoted as FC), as illustrated in Fig~\ref{fig:filter_time}. The results show that the original filter-based approach in \ac{FFMSR} exhibits lower computation time.

\begin{table}[h]
 \caption{Impact of overlapping users on \textit{Office-Arts}.}
\centering
\begin{tabular}{ll|cccc}
\toprule
\multicolumn{1}{c}{\multirow{2}{*}{Dataset}} & \multicolumn{1}{c|}{\multirow{2}{*}{Metric}} & \multicolumn{2}{c}{Recall} &  \multicolumn{2}{c}{NDCG}\\ \cmidrule(lr){3-6}
 & & @10 & @50 & @10 & @50 \\ \midrule
\multirow{2}{*}{Office} & Partial overlap & 0.1312 & 0.2069 & 0.089 & 0.1054 \\
 & No overlap & 0.1338 & 0.206 & 0.0903 & 0.106 \\ \midrule
\multirow{2}{*}{Arts} & Partial overlap & 0.1257 & 0.2377 & 0.0711 & 0.0955 \\
 & No overlap & 0.1271 & 0.2412 & 0.0712 & 0.0961 \\ \bottomrule
\end{tabular}
\label{tab:overlap}
\end{table}
\subsection{Impact of Overlapping Users}
To further validate that FFMSR does not depend on overlapping users, we conduct an additional experiment by removing all the 9,577 overlapped users in the “Office-Arts” dataset, and compared the experimental results before and after removal.
Our experimental results are presented in Table~\ref{tab:overlap}, from which we can observe that the model performs slightly better on ``non-overlapping'' datasets than on ``partial overlapping'' datasets. This phenomenon may be caused by the cold-start users in the ``partial overlapping'' datasets, since we find that overlapping users usually have relatively fewer user interactions than other non-overlapping users.
In the Office dataset, 5,895 overlapping users do not reach the average interaction length of the Office dataset, and in the Arts dataset, there are 5,361 such users.
After removing the overlapping users from the ``partial overlapping'' dataset,  the resulting ``non-overlapping'' dataset has fewer data sparsity issues within it.
Thereafter, our method can get better results on the ``non-overlapping'' datasets than the ``partial overlapping'' datasets.

%% file: 07-conclusions.tex
\section{Conclusions}
In this study, we focus on \acf{PPCDR} and propose \ac{FFMSR} to deal with the limitations of previous studies.
Specifically, to leverage the rich semantic information contained in the original text, we directly utilize the multi-layer semantic encodings produced by \ac{PLM} within the client model. During the federated training phase, the sever clusters semantic encodings from different clients to facilitate the transfer of semantic information across them. To capitalize on the synergy between different modalities, we represent items in the model using both ID and text modalities, capturing various aspects of the item's attributes. To filter out redundant noise within these modalities, we designed a gating mechanism that dynamically allocates the contribution of the ID modality and implemented a \ac{FFT}-based filtering layer to refine the text modality. Our experimental results on public datasets demonstrate \ac{FFMSR}'s ability to fully leverage information from different modalities and effectively facilitate the transfer of semantic knowledge across domains within a federated framework.

Our study has the following limitations: 1) This study is that it only leverages the encoding capabilities of \ac{LLMs}. Directly applying these models to recommendation tasks could potentially make better use of the contextual textual information from user interactions. 2) Since our approach in federated non-overlapping cross-domain recommendation depends heavily on item description text, the absence of textual descriptions or insufficiently detailed descriptions in the data domain may limit the applicability of our model. 3)  Knowledge transfer between domains with substantial semantic differences may result in negative outcomes, affecting the performance of the model. We will consider these as our future work.

%% file: sample-acmsmall.bbl

\begin{thebibliography}{67}


\ifx \showCODEN    \undefined \def \showCODEN     #1{\unskip}     \fi
\ifx \showDOI      \undefined \def \showDOI       #1{#1}\fi
\ifx \showISBNx    \undefined \def \showISBNx     #1{\unskip}     \fi
\ifx \showISBNxiii \undefined \def \showISBNxiii  #1{\unskip}     \fi
\ifx \showISSN     \undefined \def \showISSN      #1{\unskip}     \fi
\ifx \showLCCN     \undefined \def \showLCCN      #1{\unskip}     \fi
\ifx \shownote     \undefined \def \shownote      #1{#1}          \fi
\ifx \showarticletitle \undefined \def \showarticletitle #1{#1}   \fi
\ifx \showURL      \undefined \def \showURL       {\relax}        \fi
\providecommand\bibfield[2]{#2}
\providecommand\bibinfo[2]{#2}
\providecommand\natexlab[1]{#1}
\providecommand\showeprint[2][]{arXiv:#2}

\bibitem[Ai et~al\mbox{.}(2022)]%
        {ai2022fourier}
\bibfield{author}{\bibinfo{person}{Zhengyang Ai}, \bibinfo{person}{Guangjun
  Wu}, \bibinfo{person}{Binbin Li}, \bibinfo{person}{Yong Wang}, {and}
  \bibinfo{person}{Chuantong Chen}.} \bibinfo{year}{2022}\natexlab{}.
\newblock \showarticletitle{Fourier enhanced mlp with adaptive model pruning
  for efficient federated recommendation}. In
  \bibinfo{booktitle}{\emph{International Conference on Knowledge Science,
  Engineering and Management}}. Springer, \bibinfo{pages}{356--368}.
\newblock


\bibitem[Arthur et~al\mbox{.}(2007)]%
        {arthur2007k}
\bibfield{author}{\bibinfo{person}{David Arthur}, \bibinfo{person}{Sergei
  Vassilvitskii}, {et~al\mbox{.}}} \bibinfo{year}{2007}\natexlab{}.
\newblock \showarticletitle{k-means++: The advantages of careful seeding}. In
  \bibinfo{booktitle}{\emph{Soda}}, Vol.~\bibinfo{volume}{7}.
  \bibinfo{pages}{1027--1035}.
\newblock


\bibitem[Ba et~al\mbox{.}(2016)]%
        {ba2016layer}
\bibfield{author}{\bibinfo{person}{Jimmy~Lei Ba}, \bibinfo{person}{Jamie~Ryan
  Kiros}, {and} \bibinfo{person}{Geoffrey~E Hinton}.}
  \bibinfo{year}{2016}\natexlab{}.
\newblock \showarticletitle{Layer normalization}.
\newblock \bibinfo{journal}{\emph{arXiv preprint arXiv:1607.06450}}
  (\bibinfo{year}{2016}).
\newblock


\bibitem[Cai et~al\mbox{.}(2022)]%
        {cai2022privacy}
\bibfield{author}{\bibinfo{person}{Jianping Cai}, \bibinfo{person}{Yang Liu},
  \bibinfo{person}{Ximeng Liu}, \bibinfo{person}{Jiayin Li}, {and}
  \bibinfo{person}{Hongbin Zhuang}.} \bibinfo{year}{2022}\natexlab{}.
\newblock \showarticletitle{Privacy-preserving federated cross-domain social
  recommendation}. In \bibinfo{booktitle}{\emph{International Workshop on
  Trustworthy Federated Learning}}. Springer, \bibinfo{pages}{144--158}.
\newblock


\bibitem[Chen et~al\mbox{.}(2022c)]%
        {chen2022differential}
\bibfield{author}{\bibinfo{person}{Chaochao Chen}, \bibinfo{person}{Huiwen Wu},
  \bibinfo{person}{Jiajie Su}, \bibinfo{person}{Lingjuan Lyu},
  \bibinfo{person}{Xiaolin Zheng}, {and} \bibinfo{person}{Li Wang}.}
  \bibinfo{year}{2022}\natexlab{c}.
\newblock \showarticletitle{Differential private knowledge transfer for
  privacy-preserving cross-domain recommendation}. In
  \bibinfo{booktitle}{\emph{Proceedings of the ACM Web Conference 2022}}.
  \bibinfo{pages}{1455--1465}.
\newblock


\bibitem[Chen et~al\mbox{.}(2023)]%
        {chen2023win}
\bibfield{author}{\bibinfo{person}{Gaode Chen}, \bibinfo{person}{Xinghua
  Zhang}, \bibinfo{person}{Yijun Su}, \bibinfo{person}{Yantong Lai},
  \bibinfo{person}{Ji Xiang}, \bibinfo{person}{Junbo Zhang}, {and}
  \bibinfo{person}{Yu Zheng}.} \bibinfo{year}{2023}\natexlab{}.
\newblock \showarticletitle{Win-win: a privacy-preserving federated framework
  for dual-target cross-domain recommendation}. In
  \bibinfo{booktitle}{\emph{Proceedings of the AAAI Conference on Artificial
  Intelligence}}, Vol.~\bibinfo{volume}{37}. \bibinfo{pages}{4149--4156}.
\newblock


\bibitem[Chen et~al\mbox{.}(2022b)]%
        {chen2022denoising}
\bibfield{author}{\bibinfo{person}{Huiyuan Chen}, \bibinfo{person}{Yusan Lin},
  \bibinfo{person}{Menghai Pan}, \bibinfo{person}{Lan Wang},
  \bibinfo{person}{Chin-Chia~Michael Yeh}, \bibinfo{person}{Xiaoting Li},
  \bibinfo{person}{Yan Zheng}, \bibinfo{person}{Fei Wang}, {and}
  \bibinfo{person}{Hao Yang}.} \bibinfo{year}{2022}\natexlab{b}.
\newblock \showarticletitle{Denoising self-attentive sequential
  recommendation}. In \bibinfo{booktitle}{\emph{Proceedings of the 16th ACM
  Conference on Recommender Systems}}. \bibinfo{pages}{92--101}.
\newblock


\bibitem[Chen et~al\mbox{.}(2022a)]%
        {chen2022towards}
\bibfield{author}{\bibinfo{person}{Zixiang Chen}, \bibinfo{person}{Yihe Deng},
  \bibinfo{person}{Yue Wu}, \bibinfo{person}{Quanquan Gu}, {and}
  \bibinfo{person}{Yuanzhi Li}.} \bibinfo{year}{2022}\natexlab{a}.
\newblock \showarticletitle{Towards understanding mixture of experts in deep
  learning}.
\newblock \bibinfo{journal}{\emph{arXiv preprint arXiv:2208.02813}}
  (\bibinfo{year}{2022}).
\newblock


\bibitem[Cheng et~al\mbox{.}(2024)]%
        {cheng2024empowering}
\bibfield{author}{\bibinfo{person}{Mingyue Cheng}, \bibinfo{person}{Hao Zhang},
  \bibinfo{person}{Qi Liu}, \bibinfo{person}{Fajie Yuan}, \bibinfo{person}{Zhi
  Li}, \bibinfo{person}{Zhenya Huang}, \bibinfo{person}{Enhong Chen},
  \bibinfo{person}{Jun Zhou}, {and} \bibinfo{person}{Longfei Li}.}
  \bibinfo{year}{2024}\natexlab{}.
\newblock \showarticletitle{Empowering Sequential Recommendation from
  Collaborative Signals and Semantic Relatedness}.
\newblock \bibinfo{journal}{\emph{arXiv preprint arXiv:2403.07623}}
  (\bibinfo{year}{2024}).
\newblock


\bibitem[Devlin et~al\mbox{.}(2018)]%
        {devlin2018bert}
\bibfield{author}{\bibinfo{person}{Jacob Devlin}, \bibinfo{person}{Ming-Wei
  Chang}, \bibinfo{person}{Kenton Lee}, {and} \bibinfo{person}{Kristina
  Toutanova}.} \bibinfo{year}{2018}\natexlab{}.
\newblock \showarticletitle{Bert: Pre-training of deep bidirectional
  transformers for language understanding}.
\newblock \bibinfo{journal}{\emph{arXiv preprint arXiv:1810.04805}}
  (\bibinfo{year}{2018}).
\newblock


\bibitem[Ding et~al\mbox{.}(2021)]%
        {ding2021zero}
\bibfield{author}{\bibinfo{person}{Hao Ding}, \bibinfo{person}{Yifei Ma},
  \bibinfo{person}{Anoop Deoras}, \bibinfo{person}{Yuyang Wang}, {and}
  \bibinfo{person}{Hao Wang}.} \bibinfo{year}{2021}\natexlab{}.
\newblock \showarticletitle{Zero-shot recommender systems}.
\newblock \bibinfo{journal}{\emph{arXiv preprint arXiv:2105.08318}}
  (\bibinfo{year}{2021}).
\newblock


\bibitem[Du et~al\mbox{.}(2023a)]%
        {du2023distributional}
\bibfield{author}{\bibinfo{person}{Jing Du}, \bibinfo{person}{Zesheng Ye},
  \bibinfo{person}{Bin Guo}, \bibinfo{person}{Zhiwen Yu}, {and}
  \bibinfo{person}{Lina Yao}.} \bibinfo{year}{2023}\natexlab{a}.
\newblock \showarticletitle{Distributional Domain-Invariant Preference Matching
  for Cross-Domain Recommendation}. In \bibinfo{booktitle}{\emph{2023 IEEE
  International Conference on Data Mining (ICDM)}}. IEEE,
  \bibinfo{pages}{81--90}.
\newblock


\bibitem[Du et~al\mbox{.}(2023b)]%
        {du2023frequency}
\bibfield{author}{\bibinfo{person}{Xinyu Du}, \bibinfo{person}{Huanhuan Yuan},
  \bibinfo{person}{Pengpeng Zhao}, \bibinfo{person}{Jianfeng Qu},
  \bibinfo{person}{Fuzhen Zhuang}, \bibinfo{person}{Guanfeng Liu},
  \bibinfo{person}{Yanchi Liu}, {and} \bibinfo{person}{Victor~S Sheng}.}
  \bibinfo{year}{2023}\natexlab{b}.
\newblock \showarticletitle{Frequency enhanced hybrid attention network for
  sequential recommendation}. In \bibinfo{booktitle}{\emph{Proceedings of the
  46th International ACM SIGIR Conference on Research and Development in
  Information Retrieval}}. \bibinfo{pages}{78--88}.
\newblock


\bibitem[Gao et~al\mbox{.}(2022)]%
        {gao2022self}
\bibfield{author}{\bibinfo{person}{Yunjun Gao}, \bibinfo{person}{Yuntao Du},
  \bibinfo{person}{Yujia Hu}, \bibinfo{person}{Lu Chen},
  \bibinfo{person}{Xinjun Zhu}, \bibinfo{person}{Ziquan Fang}, {and}
  \bibinfo{person}{Baihua Zheng}.} \bibinfo{year}{2022}\natexlab{}.
\newblock \showarticletitle{Self-guided learning to denoise for robust
  recommendation}. In \bibinfo{booktitle}{\emph{Proceedings of the 45th
  International ACM SIGIR Conference on Research and Development in Information
  Retrieval}}. \bibinfo{pages}{1412--1422}.
\newblock


\bibitem[Guo et~al\mbox{.}(2023b)]%
        {guo2023dan}
\bibfield{author}{\bibinfo{person}{Lei Guo}, \bibinfo{person}{Hao Liu},
  \bibinfo{person}{Lei Zhu}, \bibinfo{person}{Weili Guan}, {and}
  \bibinfo{person}{Zhiyong Cheng}.} \bibinfo{year}{2023}\natexlab{b}.
\newblock \showarticletitle{DA-DAN: A Dual Adversarial Domain Adaption Network
  for Unsupervised Non-overlapping Cross-domain Recommendation}.
\newblock \bibinfo{journal}{\emph{ACM Transactions on Information Systems}}
  \bibinfo{volume}{42}, \bibinfo{number}{2} (\bibinfo{year}{2023}),
  \bibinfo{pages}{1--27}.
\newblock


\bibitem[Guo et~al\mbox{.}(2024)]%
        {guo2024prompt}
\bibfield{author}{\bibinfo{person}{Lei Guo}, \bibinfo{person}{Ziang Lu},
  \bibinfo{person}{Junliang Yu}, \bibinfo{person}{Quoc Viet~Hung Nguyen}, {and}
  \bibinfo{person}{Hongzhi Yin}.} \bibinfo{year}{2024}\natexlab{}.
\newblock \showarticletitle{Prompt-enhanced Federated Content Representation
  Learning for Cross-domain Recommendation}. In
  \bibinfo{booktitle}{\emph{Proceedings of the ACM on Web Conference 2024}}.
  \bibinfo{pages}{3139--3149}.
\newblock


\bibitem[Guo et~al\mbox{.}(2021)]%
        {guo2021gcn}
\bibfield{author}{\bibinfo{person}{Lei Guo}, \bibinfo{person}{Li Tang},
  \bibinfo{person}{Tong Chen}, \bibinfo{person}{Lei Zhu}, \bibinfo{person}{Quoc
  Viet~Hung Nguyen}, {and} \bibinfo{person}{Hongzhi Yin}.}
  \bibinfo{year}{2021}\natexlab{}.
\newblock \showarticletitle{DA-GCN: a domain-aware attentive graph convolution
  network for shared-account cross-domain sequential recommendation}. In
  \bibinfo{booktitle}{\emph{IJCAI International Joint Conference on Artificial
  Intelligence}}. International Joint Conferences on Artificial Intelligence
  Organization, \bibinfo{pages}{2483--2489}.
\newblock


\bibitem[Guo et~al\mbox{.}(2023c)]%
        {guo2023automated}
\bibfield{author}{\bibinfo{person}{Lei Guo}, \bibinfo{person}{Chunxiao Wang},
  \bibinfo{person}{Xinhua Wang}, \bibinfo{person}{Lei Zhu}, {and}
  \bibinfo{person}{Hongzhi Yin}.} \bibinfo{year}{2023}\natexlab{c}.
\newblock \showarticletitle{Automated prompting for non-overlapping
  cross-domain sequential recommendation}.
\newblock \bibinfo{journal}{\emph{arXiv preprint arXiv:2304.04218}}
  (\bibinfo{year}{2023}).
\newblock


\bibitem[Guo et~al\mbox{.}(2023a)]%
        {guo2023disentangled}
\bibfield{author}{\bibinfo{person}{Xiaobo Guo}, \bibinfo{person}{Shaoshuai Li},
  \bibinfo{person}{Naicheng Guo}, \bibinfo{person}{Jiangxia Cao},
  \bibinfo{person}{Xiaolei Liu}, \bibinfo{person}{Qiongxu Ma},
  \bibinfo{person}{Runsheng Gan}, {and} \bibinfo{person}{Yunan Zhao}.}
  \bibinfo{year}{2023}\natexlab{a}.
\newblock \showarticletitle{Disentangled representations learning for
  multi-target cross-domain recommendation}.
\newblock \bibinfo{journal}{\emph{ACM Transactions on Information Systems}}
  \bibinfo{volume}{41}, \bibinfo{number}{4} (\bibinfo{year}{2023}),
  \bibinfo{pages}{1--27}.
\newblock


\bibitem[He et~al\mbox{.}(2016)]%
        {he2016deep}
\bibfield{author}{\bibinfo{person}{Kaiming He}, \bibinfo{person}{Xiangyu
  Zhang}, \bibinfo{person}{Shaoqing Ren}, {and} \bibinfo{person}{Jian Sun}.}
  \bibinfo{year}{2016}\natexlab{}.
\newblock \showarticletitle{Deep residual learning for image recognition}. In
  \bibinfo{booktitle}{\emph{Proceedings of the IEEE conference on computer
  vision and pattern recognition}}. \bibinfo{pages}{770--778}.
\newblock


\bibitem[He et~al\mbox{.}(2020)]%
        {he2020lightgcn}
\bibfield{author}{\bibinfo{person}{Xiangnan He}, \bibinfo{person}{Kuan Deng},
  \bibinfo{person}{Xiang Wang}, \bibinfo{person}{Yan Li},
  \bibinfo{person}{Yongdong Zhang}, {and} \bibinfo{person}{Meng Wang}.}
  \bibinfo{year}{2020}\natexlab{}.
\newblock \showarticletitle{Lightgcn: Simplifying and powering graph
  convolution network for recommendation}. In
  \bibinfo{booktitle}{\emph{Proceedings of the 43rd International ACM SIGIR
  conference on research and development in Information Retrieval}}.
  \bibinfo{pages}{639--648}.
\newblock


\bibitem[Hou et~al\mbox{.}(2023)]%
        {hou2023learning}
\bibfield{author}{\bibinfo{person}{Yupeng Hou}, \bibinfo{person}{Zhankui He},
  \bibinfo{person}{Julian McAuley}, {and} \bibinfo{person}{Wayne~Xin Zhao}.}
  \bibinfo{year}{2023}\natexlab{}.
\newblock \showarticletitle{Learning vector-quantized item representation for
  transferable sequential recommenders}. In
  \bibinfo{booktitle}{\emph{Proceedings of the ACM Web Conference 2023}}.
  \bibinfo{pages}{1162--1171}.
\newblock


\bibitem[Hou et~al\mbox{.}(2022)]%
        {hou2022towards}
\bibfield{author}{\bibinfo{person}{Yupeng Hou}, \bibinfo{person}{Shanlei Mu},
  \bibinfo{person}{Wayne~Xin Zhao}, \bibinfo{person}{Yaliang Li},
  \bibinfo{person}{Bolin Ding}, {and} \bibinfo{person}{Ji-Rong Wen}.}
  \bibinfo{year}{2022}\natexlab{}.
\newblock \showarticletitle{Towards universal sequence representation learning
  for recommender systems}. In \bibinfo{booktitle}{\emph{Proceedings of the
  28th ACM SIGKDD Conference on Knowledge Discovery and Data Mining}}.
  \bibinfo{pages}{585--593}.
\newblock


\bibitem[Huang et~al\mbox{.}(2021)]%
        {huang2021whiteningbert}
\bibfield{author}{\bibinfo{person}{Junjie Huang}, \bibinfo{person}{Duyu Tang},
  \bibinfo{person}{Wanjun Zhong}, \bibinfo{person}{Shuai Lu},
  \bibinfo{person}{Linjun Shou}, \bibinfo{person}{Ming Gong},
  \bibinfo{person}{Daxin Jiang}, {and} \bibinfo{person}{Nan Duan}.}
  \bibinfo{year}{2021}\natexlab{}.
\newblock \showarticletitle{WhiteningBERT: An easy unsupervised sentence
  embedding approach}.
\newblock \bibinfo{journal}{\emph{arXiv preprint arXiv:2104.01767}}
  (\bibinfo{year}{2021}).
\newblock


\bibitem[Jin and Han(2011)]%
        {jin2011k}
\bibfield{author}{\bibinfo{person}{Xin Jin} {and} \bibinfo{person}{Jiawei
  Han}.} \bibinfo{year}{2011}\natexlab{}.
\newblock \showarticletitle{K-means clustering}.
\newblock \bibinfo{journal}{\emph{Encyclopedia of machine learning}}
  (\bibinfo{year}{2011}), \bibinfo{pages}{563--564}.
\newblock


\bibitem[Kalloori and Klingler(2021)]%
        {kalloori2021horizontal}
\bibfield{author}{\bibinfo{person}{Saikishore Kalloori} {and}
  \bibinfo{person}{Severin Klingler}.} \bibinfo{year}{2021}\natexlab{}.
\newblock \showarticletitle{Horizontal cross-silo federated recommender
  systems}. In \bibinfo{booktitle}{\emph{Proceedings of the 15th ACM Conference
  on Recommender Systems}}. \bibinfo{pages}{680--684}.
\newblock


\bibitem[Kang and McAuley(2018)]%
        {kang2018self}
\bibfield{author}{\bibinfo{person}{Wang-Cheng Kang} {and}
  \bibinfo{person}{Julian McAuley}.} \bibinfo{year}{2018}\natexlab{}.
\newblock \showarticletitle{Self-attentive sequential recommendation}. In
  \bibinfo{booktitle}{\emph{2018 IEEE international conference on data mining
  (ICDM)}}. IEEE, \bibinfo{pages}{197--206}.
\newblock


\bibitem[Li et~al\mbox{.}(2020)]%
        {li2020sentence}
\bibfield{author}{\bibinfo{person}{Bohan Li}, \bibinfo{person}{Hao Zhou},
  \bibinfo{person}{Junxian He}, \bibinfo{person}{Mingxuan Wang},
  \bibinfo{person}{Yiming Yang}, {and} \bibinfo{person}{Lei Li}.}
  \bibinfo{year}{2020}\natexlab{}.
\newblock \showarticletitle{On the sentence embeddings from pre-trained
  language models}.
\newblock \bibinfo{journal}{\emph{arXiv preprint arXiv:2011.05864}}
  (\bibinfo{year}{2020}).
\newblock


\bibitem[Li et~al\mbox{.}(2022)]%
        {li2022recguru}
\bibfield{author}{\bibinfo{person}{Chenglin Li}, \bibinfo{person}{Mingjun
  Zhao}, \bibinfo{person}{Huanming Zhang}, \bibinfo{person}{Chenyun Yu},
  \bibinfo{person}{Lei Cheng}, \bibinfo{person}{Guoqiang Shu},
  \bibinfo{person}{Beibei Kong}, {and} \bibinfo{person}{Di Niu}.}
  \bibinfo{year}{2022}\natexlab{}.
\newblock \showarticletitle{RecGURU: Adversarial learning of generalized user
  representations for cross-domain recommendation}. In
  \bibinfo{booktitle}{\emph{Proceedings of the fifteenth ACM international
  conference on web search and data mining}}. \bibinfo{pages}{571--581}.
\newblock


\bibitem[Li et~al\mbox{.}(2024)]%
        {li2024enhancing}
\bibfield{author}{\bibinfo{person}{Juanhui Li}, \bibinfo{person}{Haoyu Han},
  \bibinfo{person}{Zhikai Chen}, \bibinfo{person}{Harry Shomer},
  \bibinfo{person}{Wei Jin}, \bibinfo{person}{Amin Javari}, {and}
  \bibinfo{person}{Jiliang Tang}.} \bibinfo{year}{2024}\natexlab{}.
\newblock \showarticletitle{Enhancing ID and Text Fusion via Alternative
  Training in Session-based Recommendation}.
\newblock \bibinfo{journal}{\emph{arXiv preprint arXiv:2402.08921}}
  (\bibinfo{year}{2024}).
\newblock


\bibitem[Li et~al\mbox{.}(2023)]%
        {li2023text}
\bibfield{author}{\bibinfo{person}{Jiacheng Li}, \bibinfo{person}{Ming Wang},
  \bibinfo{person}{Jin Li}, \bibinfo{person}{Jinmiao Fu}, \bibinfo{person}{Xin
  Shen}, \bibinfo{person}{Jingbo Shang}, {and} \bibinfo{person}{Julian
  McAuley}.} \bibinfo{year}{2023}\natexlab{}.
\newblock \showarticletitle{Text is all you need: Learning language
  representations for sequential recommendation}. In
  \bibinfo{booktitle}{\emph{Proceedings of the 29th ACM SIGKDD Conference on
  Knowledge Discovery and Data Mining}}. \bibinfo{pages}{1258--1267}.
\newblock


\bibitem[Liao et~al\mbox{.}(2023)]%
        {liao2023ppgencdr}
\bibfield{author}{\bibinfo{person}{Xinting Liao}, \bibinfo{person}{Weiming
  Liu}, \bibinfo{person}{Xiaolin Zheng}, \bibinfo{person}{Binhui Yao}, {and}
  \bibinfo{person}{Chaochao Chen}.} \bibinfo{year}{2023}\natexlab{}.
\newblock \showarticletitle{Ppgencdr: A stable and robust framework for
  privacy-preserving cross-domain recommendation}. In
  \bibinfo{booktitle}{\emph{Proceedings of the AAAI Conference on Artificial
  Intelligence}}, Vol.~\bibinfo{volume}{37}. \bibinfo{pages}{4453--4461}.
\newblock


\bibitem[Liu et~al\mbox{.}(2024a)]%
        {liu2024mcrpl}
\bibfield{author}{\bibinfo{person}{Hao Liu}, \bibinfo{person}{Lei Guo},
  \bibinfo{person}{Lei Zhu}, \bibinfo{person}{Yongqiang Jiang},
  \bibinfo{person}{Min Gao}, {and} \bibinfo{person}{Hongzhi Yin}.}
  \bibinfo{year}{2024}\natexlab{a}.
\newblock \showarticletitle{MCRPL: A Pretrain, Prompt, and Fine-tune Paradigm
  for Non-overlapping Many-to-one Cross-domain Recommendation}.
\newblock \bibinfo{journal}{\emph{ACM Transactions on Information Systems}}
  \bibinfo{volume}{42}, \bibinfo{number}{4} (\bibinfo{year}{2024}),
  \bibinfo{pages}{1--24}.
\newblock


\bibitem[Liu et~al\mbox{.}(2024b)]%
        {liu2024inter}
\bibfield{author}{\bibinfo{person}{Jing Liu}, \bibinfo{person}{Lele Sun},
  \bibinfo{person}{Weizhi Nie}, \bibinfo{person}{Yuting Su},
  \bibinfo{person}{Yongdong Zhang}, {and} \bibinfo{person}{Anan Liu}.}
  \bibinfo{year}{2024}\natexlab{b}.
\newblock \showarticletitle{Inter-and Intra-Domain Potential User Preferences
  for Cross-Domain Recommendation}.
\newblock \bibinfo{journal}{\emph{IEEE Transactions on Multimedia}}
  (\bibinfo{year}{2024}).
\newblock


\bibitem[Liu et~al\mbox{.}(2023a)]%
        {liu2023differentially}
\bibfield{author}{\bibinfo{person}{Weiming Liu}, \bibinfo{person}{Xiaolin
  Zheng}, \bibinfo{person}{Chaochao Chen}, \bibinfo{person}{Mengling Hu},
  \bibinfo{person}{Xinting Liao}, \bibinfo{person}{Fan Wang},
  \bibinfo{person}{Yanchao Tan}, \bibinfo{person}{Dan Meng}, {and}
  \bibinfo{person}{Jun Wang}.} \bibinfo{year}{2023}\natexlab{a}.
\newblock \showarticletitle{Differentially private sparse mapping for
  privacy-preserving cross domain recommendation}. In
  \bibinfo{booktitle}{\emph{Proceedings of the 31st ACM International
  Conference on Multimedia}}. \bibinfo{pages}{6243--6252}.
\newblock


\bibitem[Liu et~al\mbox{.}(2023b)]%
        {liu2023joint}
\bibfield{author}{\bibinfo{person}{Weiming Liu}, \bibinfo{person}{Xiaolin
  Zheng}, \bibinfo{person}{Chaochao Chen}, \bibinfo{person}{Jiajie Su},
  \bibinfo{person}{Xinting Liao}, \bibinfo{person}{Mengling Hu}, {and}
  \bibinfo{person}{Yanchao Tan}.} \bibinfo{year}{2023}\natexlab{b}.
\newblock \showarticletitle{Joint internal multi-interest exploration and
  external domain alignment for cross domain sequential recommendation}. In
  \bibinfo{booktitle}{\emph{Proceedings of the ACM Web Conference 2023}}.
  \bibinfo{pages}{383--394}.
\newblock


\bibitem[Meihan et~al\mbox{.}(2022)]%
        {meihan2022fedcdr}
\bibfield{author}{\bibinfo{person}{Wu Meihan}, \bibinfo{person}{Li Li},
  \bibinfo{person}{Chang Tao}, \bibinfo{person}{Eric Rigall},
  \bibinfo{person}{Wang Xiaodong}, {and} \bibinfo{person}{Xu Cheng-Zhong}.}
  \bibinfo{year}{2022}\natexlab{}.
\newblock \showarticletitle{Fedcdr: federated cross-domain recommendation for
  privacy-preserving rating prediction}. In
  \bibinfo{booktitle}{\emph{Proceedings of the 31st ACM International
  Conference on Information \& Knowledge Management}}.
  \bibinfo{pages}{2179--2188}.
\newblock


\bibitem[Mu et~al\mbox{.}(2022)]%
        {mu2022id}
\bibfield{author}{\bibinfo{person}{Shanlei Mu}, \bibinfo{person}{Yupeng Hou},
  \bibinfo{person}{Wayne~Xin Zhao}, \bibinfo{person}{Yaliang Li}, {and}
  \bibinfo{person}{Bolin Ding}.} \bibinfo{year}{2022}\natexlab{}.
\newblock \showarticletitle{Id-agnostic user behavior pre-training for
  sequential recommendation}. In \bibinfo{booktitle}{\emph{China Conference on
  Information Retrieval}}. Springer, \bibinfo{pages}{16--27}.
\newblock


\bibitem[Peng et~al\mbox{.}(2023)]%
        {peng2023towards}
\bibfield{author}{\bibinfo{person}{Bo Peng}, \bibinfo{person}{Ben Burns},
  \bibinfo{person}{Ziqi Chen}, \bibinfo{person}{Srinivasan Parthasarathy},
  {and} \bibinfo{person}{Xia Ning}.} \bibinfo{year}{2023}\natexlab{}.
\newblock \showarticletitle{Towards Efficient and Effective Adaptation of Large
  Language Models for Sequential Recommendation}.
\newblock \bibinfo{journal}{\emph{arXiv preprint arXiv:2310.01612}}
  (\bibinfo{year}{2023}).
\newblock


\bibitem[Shin et~al\mbox{.}(2021)]%
        {shin2021one4all}
\bibfield{author}{\bibinfo{person}{Kyuyong Shin}, \bibinfo{person}{Hanock
  Kwak}, \bibinfo{person}{Kyung-Min Kim}, \bibinfo{person}{Minkyu Kim},
  \bibinfo{person}{Young-Jin Park}, \bibinfo{person}{Jisu Jeong}, {and}
  \bibinfo{person}{Seungjae Jung}.} \bibinfo{year}{2021}\natexlab{}.
\newblock \showarticletitle{One4all user representation for recommender systems
  in e-commerce}.
\newblock \bibinfo{journal}{\emph{arXiv preprint arXiv:2106.00573}}
  (\bibinfo{year}{2021}).
\newblock


\bibitem[Sun et~al\mbox{.}(2019)]%
        {sun2019bert4rec}
\bibfield{author}{\bibinfo{person}{Fei Sun}, \bibinfo{person}{Jun Liu},
  \bibinfo{person}{Jian Wu}, \bibinfo{person}{Changhua Pei},
  \bibinfo{person}{Xiao Lin}, \bibinfo{person}{Wenwu Ou}, {and}
  \bibinfo{person}{Peng Jiang}.} \bibinfo{year}{2019}\natexlab{}.
\newblock \showarticletitle{BERT4Rec: Sequential recommendation with
  bidirectional encoder representations from transformer}. In
  \bibinfo{booktitle}{\emph{Proceedings of the 28th ACM international
  conference on information and knowledge management}}.
  \bibinfo{pages}{1441--1450}.
\newblock


\bibitem[Sun et~al\mbox{.}(2021)]%
        {sun2021does}
\bibfield{author}{\bibinfo{person}{Yatong Sun}, \bibinfo{person}{Bin Wang},
  \bibinfo{person}{Zhu Sun}, {and} \bibinfo{person}{Xiaochun Yang}.}
  \bibinfo{year}{2021}\natexlab{}.
\newblock \showarticletitle{Does Every Data Instance Matter? Enhancing
  Sequential Recommendation by Eliminating Unreliable Data.}. In
  \bibinfo{booktitle}{\emph{IJCAI}}. \bibinfo{pages}{1579--1585}.
\newblock


\bibitem[Tian et~al\mbox{.}(2023)]%
        {tian2023privacy}
\bibfield{author}{\bibinfo{person}{Changxin Tian}, \bibinfo{person}{Yuexiang
  Xie}, \bibinfo{person}{Xu Chen}, \bibinfo{person}{Yaliang Li}, {and}
  \bibinfo{person}{Wayne~Xin Zhao}.} \bibinfo{year}{2023}\natexlab{}.
\newblock \showarticletitle{Privacy-Preserving Cross-Domain Recommendation with
  Federated Graph Learning}.
\newblock \bibinfo{journal}{\emph{ACM Transactions on Information Systems}}
  (\bibinfo{year}{2023}).
\newblock


\bibitem[Vaswani et~al\mbox{.}(2017)]%
        {vaswani2017attention}
\bibfield{author}{\bibinfo{person}{Ashish Vaswani}, \bibinfo{person}{Noam
  Shazeer}, \bibinfo{person}{Niki Parmar}, \bibinfo{person}{Jakob Uszkoreit},
  \bibinfo{person}{Llion Jones}, \bibinfo{person}{Aidan~N Gomez},
  \bibinfo{person}{{\L}ukasz Kaiser}, {and} \bibinfo{person}{Illia
  Polosukhin}.} \bibinfo{year}{2017}\natexlab{}.
\newblock \showarticletitle{Attention is all you need}.
\newblock \bibinfo{journal}{\emph{Advances in neural information processing
  systems}}  \bibinfo{volume}{30} (\bibinfo{year}{2017}).
\newblock


\bibitem[Voigt and Von~dem Bussche(2017)]%
        {voigt2017eu}
\bibfield{author}{\bibinfo{person}{Paul Voigt} {and} \bibinfo{person}{Axel
  Von~dem Bussche}.} \bibinfo{year}{2017}\natexlab{}.
\newblock \showarticletitle{The eu general data protection regulation (gdpr)}.
\newblock \bibinfo{journal}{\emph{A Practical Guide, 1st Ed., Cham: Springer
  International Publishing}} \bibinfo{volume}{10}, \bibinfo{number}{3152676}
  (\bibinfo{year}{2017}), \bibinfo{pages}{10--5555}.
\newblock


\bibitem[Wan et~al\mbox{.}(2023)]%
        {wan2023fedpdd}
\bibfield{author}{\bibinfo{person}{Sheng Wan}, \bibinfo{person}{Dashan Gao},
  \bibinfo{person}{Hanlin Gu}, {and} \bibinfo{person}{Daning Hu}.}
  \bibinfo{year}{2023}\natexlab{}.
\newblock \showarticletitle{FedPDD: A Privacy-preserving Double Distillation
  Framework for Cross-silo Federated Recommendation}. In
  \bibinfo{booktitle}{\emph{2023 International Joint Conference on Neural
  Networks (IJCNN)}}. IEEE, \bibinfo{pages}{1--8}.
\newblock


\bibitem[Wang et~al\mbox{.}(2024a)]%
        {wang2024privacy}
\bibfield{author}{\bibinfo{person}{Li Wang}, \bibinfo{person}{Lei Sang},
  \bibinfo{person}{Quangui Zhang}, \bibinfo{person}{Qiang Wu}, {and}
  \bibinfo{person}{Min Xu}.} \bibinfo{year}{2024}\natexlab{a}.
\newblock \showarticletitle{A Privacy-Preserving Framework with Multi-Modal
  Data for Cross-Domain Recommendation}.
\newblock \bibinfo{journal}{\emph{arXiv preprint arXiv:2403.03600}}
  (\bibinfo{year}{2024}).
\newblock


\bibitem[Wang et~al\mbox{.}(2024b)]%
        {wang2024aligned}
\bibfield{author}{\bibinfo{person}{Shuhan Wang}, \bibinfo{person}{Bin Shen},
  \bibinfo{person}{Xu Min}, \bibinfo{person}{Yong He}, \bibinfo{person}{Xiaolu
  Zhang}, \bibinfo{person}{Liang Zhang}, \bibinfo{person}{Jun Zhou}, {and}
  \bibinfo{person}{Linjian Mo}.} \bibinfo{year}{2024}\natexlab{b}.
\newblock \showarticletitle{Aligned Side Information Fusion Method for
  Sequential Recommendation}. In \bibinfo{booktitle}{\emph{Companion
  Proceedings of the ACM on Web Conference 2024}}. \bibinfo{pages}{112--120}.
\newblock


\bibitem[Wang et~al\mbox{.}(2021)]%
        {wang2021denoising}
\bibfield{author}{\bibinfo{person}{Wenjie Wang}, \bibinfo{person}{Fuli Feng},
  \bibinfo{person}{Xiangnan He}, \bibinfo{person}{Liqiang Nie}, {and}
  \bibinfo{person}{Tat-Seng Chua}.} \bibinfo{year}{2021}\natexlab{}.
\newblock \showarticletitle{Denoising implicit feedback for recommendation}. In
  \bibinfo{booktitle}{\emph{Proceedings of the 14th ACM international
  conference on web search and data mining}}. \bibinfo{pages}{373--381}.
\newblock


\bibitem[Wang et~al\mbox{.}(2019)]%
        {wang2019neural}
\bibfield{author}{\bibinfo{person}{Xiang Wang}, \bibinfo{person}{Xiangnan He},
  \bibinfo{person}{Meng Wang}, \bibinfo{person}{Fuli Feng}, {and}
  \bibinfo{person}{Tat-Seng Chua}.} \bibinfo{year}{2019}\natexlab{}.
\newblock \showarticletitle{Neural graph collaborative filtering}. In
  \bibinfo{booktitle}{\emph{Proceedings of the 42nd international ACM SIGIR
  conference on Research and development in Information Retrieval}}.
  \bibinfo{pages}{165--174}.
\newblock


\bibitem[Xie et~al\mbox{.}(2022)]%
        {xie2022contrastive}
\bibfield{author}{\bibinfo{person}{Ruobing Xie}, \bibinfo{person}{Qi Liu},
  \bibinfo{person}{Liangdong Wang}, \bibinfo{person}{Shukai Liu},
  \bibinfo{person}{Bo Zhang}, {and} \bibinfo{person}{Leyu Lin}.}
  \bibinfo{year}{2022}\natexlab{}.
\newblock \showarticletitle{Contrastive cross-domain recommendation in
  matching}. In \bibinfo{booktitle}{\emph{Proceedings of the 28th ACM SIGKDD
  Conference on Knowledge Discovery and Data Mining}}.
  \bibinfo{pages}{4226--4236}.
\newblock


\bibitem[Xu et~al\mbox{.}(2024)]%
        {xu2024sequence}
\bibfield{author}{\bibinfo{person}{Lanling Xu}, \bibinfo{person}{Zhen Tian},
  \bibinfo{person}{Bingqian Li}, \bibinfo{person}{Junjie Zhang},
  \bibinfo{person}{Daoyuan Wang}, \bibinfo{person}{Hongyu Wang},
  \bibinfo{person}{Jinpeng Wang}, \bibinfo{person}{Sheng Chen}, {and}
  \bibinfo{person}{Wayne~Xin Zhao}.} \bibinfo{year}{2024}\natexlab{}.
\newblock \showarticletitle{Sequence-level Semantic Representation Fusion for
  Recommender Systems}. In \bibinfo{booktitle}{\emph{Proceedings of the 33rd
  ACM International Conference on Information and Knowledge Management}}.
  \bibinfo{pages}{5015--5022}.
\newblock


\bibitem[Yan et~al\mbox{.}(2022)]%
        {yan2022fedcdr}
\bibfield{author}{\bibinfo{person}{Dengcheng Yan}, \bibinfo{person}{Yuchuan
  Zhao}, \bibinfo{person}{Zhongxiu Yang}, \bibinfo{person}{Ying Jin}, {and}
  \bibinfo{person}{Yiwen Zhang}.} \bibinfo{year}{2022}\natexlab{}.
\newblock \showarticletitle{FedCDR: Privacy-preserving federated cross-domain
  recommendation}.
\newblock \bibinfo{journal}{\emph{Digital Communications and Networks}}
  \bibinfo{volume}{8}, \bibinfo{number}{4} (\bibinfo{year}{2022}),
  \bibinfo{pages}{552--560}.
\newblock


\bibitem[Yuan et~al\mbox{.}(2023)]%
        {yuan2023go}
\bibfield{author}{\bibinfo{person}{Zheng Yuan}, \bibinfo{person}{Fajie Yuan},
  \bibinfo{person}{Yu Song}, \bibinfo{person}{Youhua Li},
  \bibinfo{person}{Junchen Fu}, \bibinfo{person}{Fei Yang},
  \bibinfo{person}{Yunzhu Pan}, {and} \bibinfo{person}{Yongxin Ni}.}
  \bibinfo{year}{2023}\natexlab{}.
\newblock \showarticletitle{Where to go next for recommender systems? id-vs.
  modality-based recommender models revisited}. In
  \bibinfo{booktitle}{\emph{Proceedings of the 46th International ACM SIGIR
  Conference on Research and Development in Information Retrieval}}.
  \bibinfo{pages}{2639--2649}.
\newblock


\bibitem[Zhang et~al\mbox{.}(2024b)]%
        {zhang2024feddcsr}
\bibfield{author}{\bibinfo{person}{Hongyu Zhang}, \bibinfo{person}{Dongyi
  Zheng}, \bibinfo{person}{Xu Yang}, \bibinfo{person}{Jiyuan Feng}, {and}
  \bibinfo{person}{Qing Liao}.} \bibinfo{year}{2024}\natexlab{b}.
\newblock \showarticletitle{FedDCSR: Federated cross-domain sequential
  recommendation via disentangled representation learning}. In
  \bibinfo{booktitle}{\emph{Proceedings of the 2024 SIAM International
  Conference on Data Mining (SDM)}}. SIAM, \bibinfo{pages}{535--543}.
\newblock


\bibitem[Zhang et~al\mbox{.}(2024c)]%
        {zhang2024fedhcdr}
\bibfield{author}{\bibinfo{person}{Hongyu Zhang}, \bibinfo{person}{Dongyi
  Zheng}, \bibinfo{person}{Lin Zhong}, \bibinfo{person}{Xu Yang},
  \bibinfo{person}{Jiyuan Feng}, \bibinfo{person}{Yunqing Feng}, {and}
  \bibinfo{person}{Qing Liao}.} \bibinfo{year}{2024}\natexlab{c}.
\newblock \showarticletitle{FedHCDR: Federated Cross-Domain Recommendation with
  Hypergraph Signal Decoupling}. In \bibinfo{booktitle}{\emph{Joint European
  Conference on Machine Learning and Knowledge Discovery in Databases}}.
  Springer, \bibinfo{pages}{350--366}.
\newblock


\bibitem[Zhang and Jiang(2021)]%
        {zhang2021vertical}
\bibfield{author}{\bibinfo{person}{JianFei Zhang} {and} \bibinfo{person}{YuChen
  Jiang}.} \bibinfo{year}{2021}\natexlab{}.
\newblock \showarticletitle{A vertical federation recommendation method based
  on clustering and latent factor model}. In \bibinfo{booktitle}{\emph{2021
  International Conference on Electronic Information Engineering and Computer
  Science (EIECS)}}. IEEE, \bibinfo{pages}{362--366}.
\newblock


\bibitem[Zhang et~al\mbox{.}(2024d)]%
        {zhang2024id}
\bibfield{author}{\bibinfo{person}{Lingzi Zhang}, \bibinfo{person}{Xin Zhou},
  \bibinfo{person}{Zhiwei Zeng}, {and} \bibinfo{person}{Zhiqi Shen}.}
  \bibinfo{year}{2024}\natexlab{d}.
\newblock \showarticletitle{Are id embeddings necessary? whitening pre-trained
  text embeddings for effective sequential recommendation}.
\newblock \bibinfo{journal}{\emph{arXiv preprint arXiv:2402.10602}}
  (\bibinfo{year}{2024}).
\newblock


\bibitem[Zhang et~al\mbox{.}(2019)]%
        {zhang2019feature}
\bibfield{author}{\bibinfo{person}{Tingting Zhang}, \bibinfo{person}{Pengpeng
  Zhao}, \bibinfo{person}{Yanchi Liu}, \bibinfo{person}{Victor~S Sheng},
  \bibinfo{person}{Jiajie Xu}, \bibinfo{person}{Deqing Wang},
  \bibinfo{person}{Guanfeng Liu}, \bibinfo{person}{Xiaofang Zhou},
  {et~al\mbox{.}}} \bibinfo{year}{2019}\natexlab{}.
\newblock \showarticletitle{Feature-level Deeper Self-Attention Network for
  Sequential Recommendation.}. In \bibinfo{booktitle}{\emph{IJCAI}}.
  \bibinfo{pages}{4320--4326}.
\newblock


\bibitem[Zhang et~al\mbox{.}(2024a)]%
        {zhang2024disentangling}
\bibfield{author}{\bibinfo{person}{Xiaokun Zhang}, \bibinfo{person}{Bo Xu},
  \bibinfo{person}{Zhaochun Ren}, \bibinfo{person}{Xiaochen Wang},
  \bibinfo{person}{Hongfei Lin}, {and} \bibinfo{person}{Fenglong Ma}.}
  \bibinfo{year}{2024}\natexlab{a}.
\newblock \showarticletitle{Disentangling id and modality effects for
  session-based recommendation}. In \bibinfo{booktitle}{\emph{Proceedings of
  the 47th International ACM SIGIR Conference on Research and Development in
  Information Retrieval}}. \bibinfo{pages}{1883--1892}.
\newblock


\bibitem[Zhang et~al\mbox{.}(2022)]%
        {zhang2022attention}
\bibfield{author}{\bibinfo{person}{Yichi Zhang}, \bibinfo{person}{Guisheng
  Yin}, \bibinfo{person}{Hongbin Dong}, {and} \bibinfo{person}{Liguo Zhang}.}
  \bibinfo{year}{2022}\natexlab{}.
\newblock \showarticletitle{Attention-based frequency-aware multi-scale network
  for sequential recommendation}.
\newblock \bibinfo{journal}{\emph{Applied Soft Computing}}
  \bibinfo{volume}{127} (\bibinfo{year}{2022}), \bibinfo{pages}{109349}.
\newblock


\bibitem[Zhang et~al\mbox{.}(2023a)]%
        {zhang2023contrastivet}
\bibfield{author}{\bibinfo{person}{Yichi Zhang}, \bibinfo{person}{Guisheng
  Yin}, {and} \bibinfo{person}{Yuxin Dong}.} \bibinfo{year}{2023}\natexlab{a}.
\newblock \showarticletitle{Contrastive learning with frequency-domain interest
  trends for sequential recommendation}. In
  \bibinfo{booktitle}{\emph{Proceedings of the 17th ACM Conference on
  Recommender Systems}}. \bibinfo{pages}{141--150}.
\newblock


\bibitem[Zhang et~al\mbox{.}(2023b)]%
        {zhang2023contrastive}
\bibfield{author}{\bibinfo{person}{Yichi Zhang}, \bibinfo{person}{Guisheng
  Yin}, \bibinfo{person}{Yuxin Dong}, {and} \bibinfo{person}{Liguo Zhang}.}
  \bibinfo{year}{2023}\natexlab{b}.
\newblock \showarticletitle{Contrastive Learning with Frequency Domain for
  Sequential Recommendation}.
\newblock \bibinfo{journal}{\emph{Applied Soft Computing}}
  \bibinfo{volume}{144} (\bibinfo{year}{2023}), \bibinfo{pages}{110481}.
\newblock


\bibitem[Zhao et~al\mbox{.}(2023)]%
        {zhao2023cross}
\bibfield{author}{\bibinfo{person}{Chuang Zhao}, \bibinfo{person}{Hongke Zhao},
  \bibinfo{person}{Ming He}, \bibinfo{person}{Jian Zhang}, {and}
  \bibinfo{person}{Jianping Fan}.} \bibinfo{year}{2023}\natexlab{}.
\newblock \showarticletitle{Cross-domain recommendation via user interest
  alignment}. In \bibinfo{booktitle}{\emph{Proceedings of the ACM Web
  Conference 2023}}. \bibinfo{pages}{887--896}.
\newblock


\bibitem[Zhao et~al\mbox{.}(2024)]%
        {zhao2024personalized}
\bibfield{author}{\bibinfo{person}{Peng Zhao}, \bibinfo{person}{Yuanyang Jin},
  \bibinfo{person}{Xuebin Ren}, {and} \bibinfo{person}{Yanan Li}.}
  \bibinfo{year}{2024}\natexlab{}.
\newblock \showarticletitle{A personalized cross-domain recommendation with
  federated meta learning}.
\newblock \bibinfo{journal}{\emph{Multimedia Tools and Applications}}
  (\bibinfo{year}{2024}), \bibinfo{pages}{1--16}.
\newblock


\bibitem[Zhou et~al\mbox{.}(2020)]%
        {zhou2020s3}
\bibfield{author}{\bibinfo{person}{Kun Zhou}, \bibinfo{person}{Hui Wang},
  \bibinfo{person}{Wayne~Xin Zhao}, \bibinfo{person}{Yutao Zhu},
  \bibinfo{person}{Sirui Wang}, \bibinfo{person}{Fuzheng Zhang},
  \bibinfo{person}{Zhongyuan Wang}, {and} \bibinfo{person}{Ji-Rong Wen}.}
  \bibinfo{year}{2020}\natexlab{}.
\newblock \showarticletitle{S3-rec: Self-supervised learning for sequential
  recommendation with mutual information maximization}. In
  \bibinfo{booktitle}{\emph{Proceedings of the 29th ACM international
  conference on information \& knowledge management}}.
  \bibinfo{pages}{1893--1902}.
\newblock


\bibitem[Zhou et~al\mbox{.}(2022)]%
        {zhou2022filter}
\bibfield{author}{\bibinfo{person}{Kun Zhou}, \bibinfo{person}{Hui Yu},
  \bibinfo{person}{Wayne~Xin Zhao}, {and} \bibinfo{person}{Ji-Rong Wen}.}
  \bibinfo{year}{2022}\natexlab{}.
\newblock \showarticletitle{Filter-enhanced MLP is all you need for sequential
  recommendation}. In \bibinfo{booktitle}{\emph{Proceedings of the ACM web
  conference 2022}}. \bibinfo{pages}{2388--2399}.
\newblock


\end{thebibliography}
